\documentclass[preprint,12pt]{aastex62}

\shorttitle{L1157 Binary}
\shortauthors{Tobin et al.}

\newcommand{\nthp}{\mbox{N$_2$H$^+$}}

\newcommand{\nht}{\mbox{NH$_3$}}

\newcommand{\cateo}{\mbox{C$^{18}$O}}

\newcommand{\kmspc}{\mbox{km s$^{-1}$ pc$^{-1}$ }}
\newcommand{\kms}{\mbox{km s$^{-1}$}}

\newcommand{\lsun}{\mbox{L$_{\sun}$}}
\newcommand{\msun}{\mbox{M$_{\sun}$}}
\newcommand{\mearth}{\mbox{M$_{\oplus}$}}

\begin{document}

\title{A 16~au Binary in the Class 0 Protostar L1157~MMS}
\author[0000-0002-6195-0152]{John J. Tobin}
\affiliation{National Radio Astronomy Observatory, 520 Edgemont Rd., Charlottesville,VA 22903, USA}
\author[0000-0002-5216-8062]{Erin G. Cox}
\affiliation{Center
for Interdisciplinary Exploration and Research in Astronomy, 1800 Sherman Rd., Evanston, IL 60202, USA}
\author[0000-0002-4540-6587]{Leslie W. Looney}
\affiliation{Department of Astronomy, University of Illinois, Urbana, IL 61801, USA}

\begin{abstract}
We present VLA observations toward the Class 0 protostar L1157 MMS at 6.8~mm and 9~mm with
a resolution of $\sim$0\farcs04 (14~au). We detect two sources within L1157 MMS and 
interpret these sources as a binary protostar with a separation of $\sim$16~au.
The material directly surrounding the binary system within the 
inner 50~au radius of the system has an estimated mass of 0.11~\msun, calculated
from the observed dust emission. We interpret the observed binary system
in the context of previous 
observations of its flattened envelope
structure, low rates of envelope rotation from 5000 to 200~au scales, and
an ordered, poloidal magnetic field aligned with the outflow. Thus, 
L1157~MMS is a prototype system for magnetically-regulated collapse and the
presence of a compact binary within L1157 MMS demonstrates that multiple star formation can still occur within
envelopes that likely have dynamically important magnetic fields.
\end{abstract}

\section{Introduction}

Star formation occurs within dense clouds of gas and dust
that are collapsing under their own gravity, leading to the formation
of one or more stars. In addition to gravity, the dynamics of 
collapse is expected to be influenced by turbulence, angular momentum, and magnetic
fields \citep{li2014}. A disk is expected to form around the nascent protostar(s)
during collapse via conservation of angular momentum,
and this disk can lead to both the formation
of companion stars and/or planets. Our collective understanding of star formation
is the result of a combination of survey-mode studies that can answer
specific questions due to the large samples, and case-studies of individual
systems that clearly display particular aspects of the star formation process
by their nature of proximity, brightness, and morphological simplicity. One
such system that exemplifies many aspects of the star formation process is
the protostar system L1157 MMS.

L1157 MMS is a Class 0 protostar \citep{gueth1996}, the youngest class of protostar with a 
dense infalling envelope of gas and dust \citep{andre1993}, located at a distance of $\sim$352~pc 
\citep{zucker2019}. L1157 MMS is an isolated system within the Cepheus Flare
clouds, having no neighboring protostars within 1~pc. Thus, this system is evolving
without significant influence of nearby outflows or any apparent interaction
with past star formation events, making it an ideal source for star formation
case studies. 

This system was initially best known for its 
outflow \citep[e.g.,][]{bachiller1993,gueth1996,bachiller2001}, but  
an extended (R $\sim$10$^4$~au), flattened envelope was found
in 8~\micron\ extinction of Galactic background emission \citep{looney2007}
whose major axis is nearly orthogonal to the outflow.
The envelope also has \nthp\ and \nht\ emission that closely maps to the
8~\micron\ extinction \citep{chiang2010,tobin2011}. Flattened continuum emission
from the inner envelope was also observed at 1.3~mm with
the Submillimeter Array \citep{tobin2013}. Within this flattened envelope, 
magnetic fields have been observed via dust polarization showing a poloidal
field geometry, aligned with the outflow and perpendicular to the 
flattened envelope on both 1000~au and 5000~au scales \citep{chapman2013,stephens2013}.
Thus, L1157 MMS outwardly has the appearance that may reflect a prototypical
case of magnetically-regulated star formation, in which magnetic fields promote
the formation of flattened envelopes via collapse along field lines and
ambipolar diffusion of magnetic flux to allow infall orthogonal to the magnetic
field lines, enabling formation of the protostar \citep[e.g.,][]{galli1993}.

The gas kinematics of the flattened envelope only show very small velocity gradients
on scales $>$2000~au, with a measured value of 0.9~\kmspc\ \citep{tobin2011}.
Furthermore, higher-resolution observations of the inner envelope ($\sim$500~au scales)
in \cateo\ by \citet{gaudel2020}, where the kinematics are
expected to be dominated by rotation, show little evidence for rotation that
reflects angular momentum conservation, let alone a Keplerian disk.
This is further evidence that the collapse of L1157 MMS may be dominated by 
magnetic fields, because magnetic braking can remove angular momentum 
from infalling material, thereby suppressing any increases in rotation 
velocity expected from conservation of angular momentum \citep{basu1995,allen2003}. 
Thus, L1157 MMS presents an ideal prototype
for studying the disk structure within an envelope where magnetic fields
are dynamically important.

\citet{chiang2012} studied L1157 MMS with the highest angular resolution
possible at millimeter wavelengths ($\sim$0\farcs3) and did not resolve the disk in continuum 
emission, finding that any disk must be smaller than 40~au in radius. The northern location
currently prohibits higher resolution observations at millimeter wavelengths,
but the NSF's Karl G. Jansky Very Large Array (VLA) can provide 
resolution as fine as 0\farcs04 at a wavelength of $\sim$6.8~mm, enabling 
$<$~20~au resolution toward this protostar, the highest currently possible.
\citet{tobin2013} observed this source during the early-science phase of the
upgraded VLA, finding compact emission with a slight extension orthogonal to the
outflow, but no obvious signs of a disk or multiplicity. 

We now follow-up
those observations using the full sensitivity of the VLA at 6.8~mm (Q-band) and 9~mm (Ka-band)
to better characterize the small-scale structure toward this protostar.
The paper is organized as follows: the observations are described
in Section 2, the observational results are presented in Section 3, 
the results are discussed in Section 4, and we present our conclusions 
in Section 5.

\section{Observations and Data Reduction}

We observed L1157 MMS with the VLA, located on the plains of San Agustin in New Mexico, USA,
in 2015 during A-configuration as part of VLA program 15A-370. 
Our observations of L1157 MMS were conducted in four executions (dates listed in Table \ref{vla_obs}),
and during each execution it was observed at both Q and Ka bands with a
time on source of $\sim$30~min in each band. 
Pointing was updated once an hour when observing L1157 MMS and prior to observing
the flux density (3C48) and bandpass (3C84) calibrators. Fast switching was utilized
to compensate for rapid atmospheric phase variations with a total calibration
cycle time (calibrator, source, and calibrator) of 170~s in Q and Ka-bands;
J2006+6424 was used as the complex gain calibrator.
The correlator was used in 3-bit mode for both bands, providing 8~GHz of bandwidth
in each band.
The Ka-band basebands were centered at 29 GHz and 37 GHz (10.3 and 8.1~mm), while the Q-band
basebands were centered at 42~GHz and 46 GHz (7.1 and 6.5~mm); the first observation centered
the Q-band basebands at 41~GHz and 45~GHz (attempting to avoid the sensitivity decrease from 47 to 48~GHz), but this yielded higher noise, so
the aforementioned tunings were used for the subsequent observations. 
The Q-band observations sampled uv-distances of $\sim$100 to 5000~k$\lambda$, and the Ka-band 
observations sampled $\sim$74 to 4000~k$\lambda$. The uv ranges are approximate given that they correspond to the band
centers.

The data were reduced using the VLA calibration pipeline version 2020.1 in CASA 6.1.2.
We used 
multiple pipeline runs for each observation. The first pipeline run was used to identify 
misbehaving antennas, and we flagged these data and re-ran the pipeline. Then, 
following the 
pipeline run, we performed any additional flagging using the final gain calibration tables
and applied the flags using the CASA task \textit{applycal} with \textit{mode=`flagonlystrict'}. We 
then inspected raw target data and flagged any additional data that were outliers. 
The absolute flux calibration accuracy is expected to be $\sim$10\%.

We also
conducted polarization calibration using the pipeline calibration as a starting point, following the published CASA guide\footnote{https://casaguides.nrao.edu/index.php/ \newline
Polarization\_Calibration\_based\_on\_CASA\_pipeline\_standard\_reduction:\_The\_radio\_galaxy\_3C75}. We used 3C84 as the unpolarized leakage calibrator, using \textit{poltype=`Df'} in the CASA task \textit{polcal}.
The calibrator 3C48 was used for the cross-hand delay calibration and the polarization angle calibration; when computing the polarization angle, we used a \textit{uvrange} selection of 0 to 200~k$\lambda$ to avoid errors in angle calibration due to 3C48 being well-resolved at this frequency and angular resolution.

The S/N of the data were high enough to allow for self-calibration to be attempted. We used
the combined Ka and Q band data and created the model for self-calibration using the 
data imaged with a \textit{robust} parameter of 2.0. We computed phase-only solutions per antenna,
for all the spectral windows combined from both bands,
with a solution interval that spanned the entire length of each observation. The corrections
computed for each observation have the effect of reducing systematic position shifts between
observations that result from phase transfer errors. These phase transfer errors 
can result from small antenna
position errors \citep{brogan2018}. Subsequent self-calibration attempts with shorter 
solution intervals did not increase the S/N and our final data only have the per-observation
solutions applied. 
We use the self-calibrated data for all images and measurements except
where specified.

We imaged the data with CASA 6.1.2 using the task \textit{tclean}. We made several images
using Briggs weighting with different values for the \textit{robust} parameter to emphasize
structure at different scales in the data. We also made images of different frequency ranges to 
assess the robustness of the detected structure.

\section{Results}
\begin{figure}
\begin{center}
\includegraphics[scale=0.475]{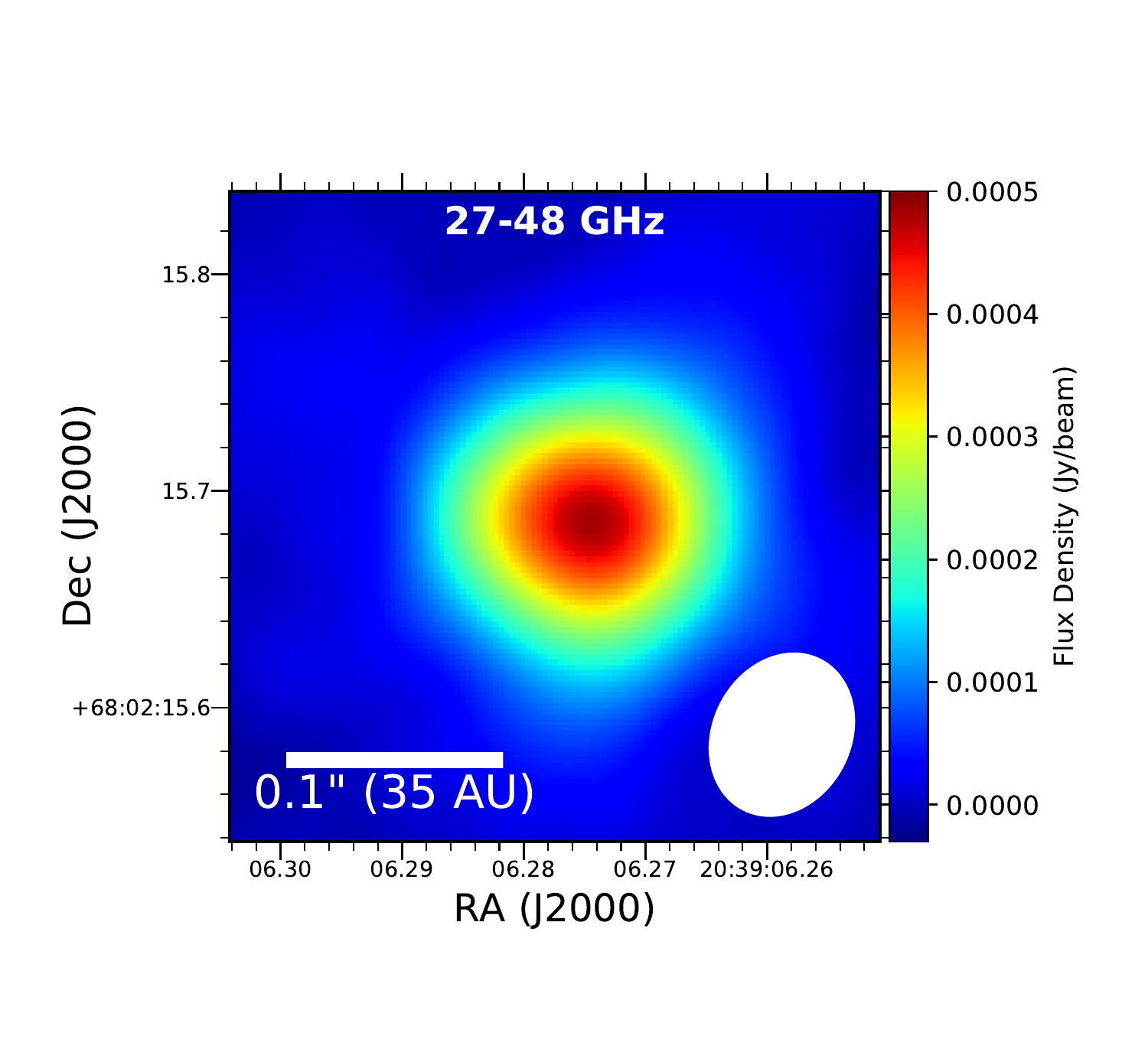}
\includegraphics[scale=0.475]{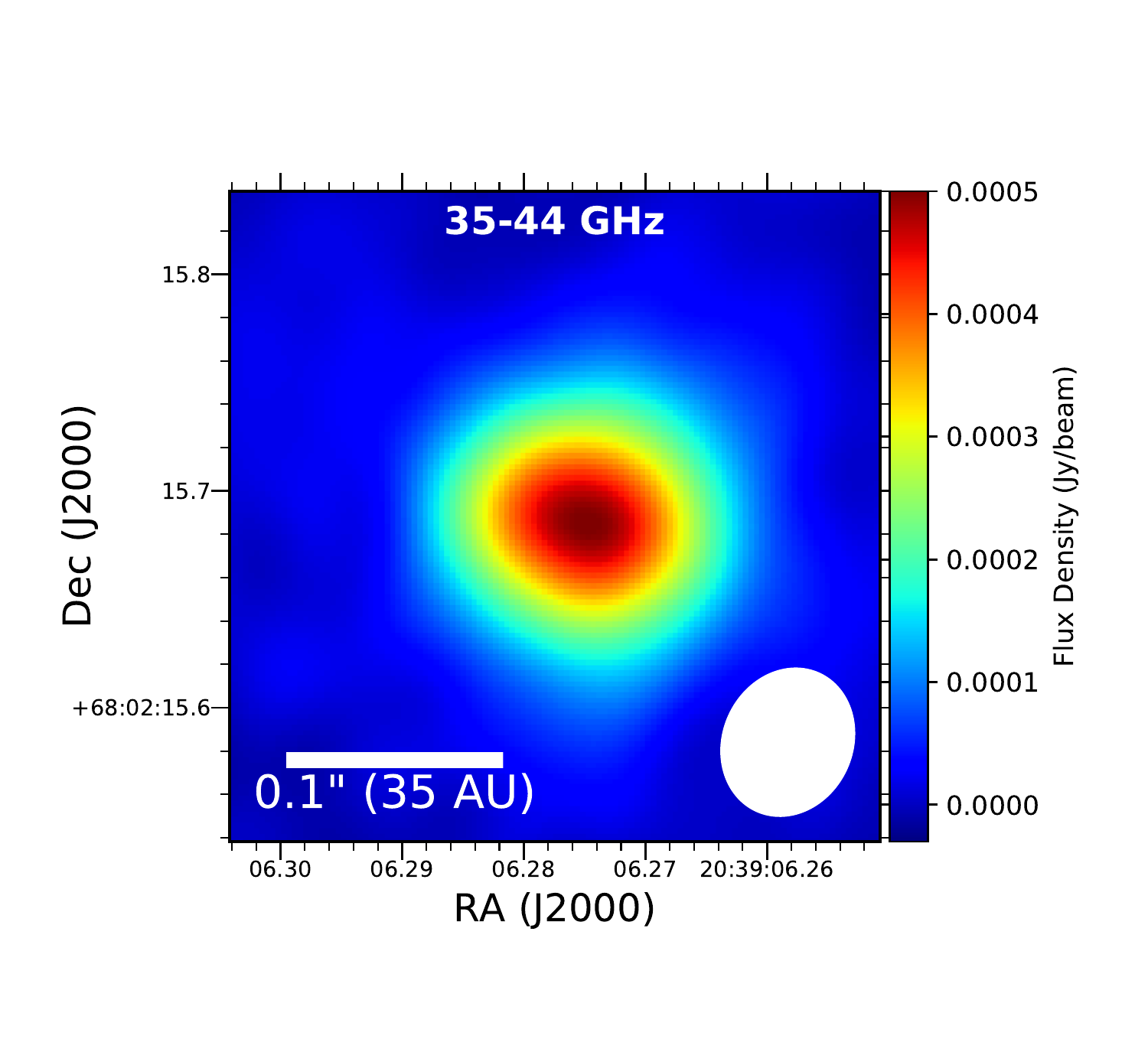}

\end{center}
\caption{Images of L1157 MMS 
The sources are clearly resolved in both frequency ranges with respect to the beam,
but details of the structure cannot be discerned at this resolution. 
The main impact of self-calibration on images at this resolution is
a higher peak intensity in both images as compared to the non-self-calibrated data.
The values for the beam sizes in each image are provided in Table \ref{image-parameters}.
}
\label{continuum-lowres}
\end{figure}

L1157 MMS is clearly detected in our Q and Ka-band imaging in A-configuration. 
We show images of L1157 MMS in Figure \ref{continuum-lowres} generated 
with robust=2 from the combined Q and Ka-band data,
providing the best sensitivity. The source
has a similar morphology when imaged with the full bandwidth vs. only the 35 to 44~GHz 
range, also shown in Figure \ref{continuum-lowres}. The source is resolved
compared to the synthesized beam, and the peak intensity in the combined
Q and Ka-band image is 0.47~mJy~beam$^{-1}$, while the integrated flux density
is $\sim$1.1~mJy, further demonstrating that the source is resolved.

We made images for the full range of the combined Q and Ka-band data and the 35 to 44~GHz range
using $robust=-0.25$ to add more weight to the longer baselines during imaging, resulting
in higher resolution images. These two higher resolution images in Figure \ref{continuum-highres} 
show that the continuum emission is resolved into two peaks. The peak intensity of the western
peak (0.27~mJy~beam$^{-1}$) is about 1.3$\times$ greater than the eastern peak (0.20~mJy~beam$^{-1}$).
We denote the western peak as component A and the eastern peak as component B.
These resolved peaks have a position angle comparable
to the marginally-resolved 7.3~mm image of L1157 MMS from \citet{tobin2013}.
 The new data (Q and Ka-bands combined) have
a factor of 2.7 lower noise and higher angular resolution than the previous
observations. The improved S/N is a result of
the wider bandwidth and longer integration time for the new data. We also show a comparison
of the images generated with self-calibration applied and those without self-calibration
applied in Figure \ref{continuum-highres}. This demonstrates that the two sources
are detected and resolved whether self-calibration is applied or not, but the significance 
of the eastern component is increased by $>$2$\sigma$ in the self-calibrated images.

The positions of the continuum sources are listed in Table \ref{sources},
along with the values from the a reprocessing and analysis of the 2012 data (see Appendix), 
and the beams and RMS noises in each image used for the Figures
are given in Table \ref{image-parameters}. The overall source position is measured using the 
robust=2.0 image from the combined Q and Ka-band data, while the positions of
the individual resolve sources are measured from the robust=-0.25 image using the combined
Q and Ka-band data. The continuum positions were fitted using the 
CASA task \textit{imfit}, and when fitting the positions of the two 
continuum sources, we fixed the size of the Gaussians to the size of the
synthesized beam. 

The time baseline of $\sim$2.6~yr from the observations in 2012 to
the new observations in 2015 enables us to calculate the proper motion of L1157 MMS as a whole (not the
individual components A and B). 
We calculate a proper motion of 23.9$\pm$2.1~mas~yr$^{-1}$ for 
Right Ascension and 2.3$\pm$2.0~mas~yr$^{-1}$  for Declination using the coordinates from each epoch
of observation.

\subsection{Binary Protostar}
The double-peaked morphology of the image toward L1157 MMS 
suggests that L1157 MMS is a proto-binary system with a separation of 0\farcs044$\pm$0.002 ($\sim$16~au).
This is one of the closest Class 0 proto-binary systems known. Its angular and
physical separation is smaller than the most compact multiple systems 
observed by \citet{tobin2016,tobin2021} in the VLA and ALMA Nascent Disk 
and Multiplicity (VANDAM) Surveys. We examined the data in several ways to
ensure the robustness of our detection detection of two sources within L1157~MMS, and we
outline the additional analysis in the Appendix.

\subsection{Polarization Upper Limits}

As part of our high-resolution observations, we also 
examined the polarimetry toward L1157 MMS in Q-band. We constructed Stokes
IQUV cubes using \textit{Natural} weighting and with a uv-taper at 1000~k$\lambda$ 
to focus on larger angular scales. However, on the size scales probed in those images
(0\farcs06 and 0\farcs14), we do not detect significant Stokes Q or U emission. Thus, we can 
only provide 3$\sigma$ upper limits on the total polarized peak intensity of 0.1 mJy beam$^{-1}$ 
and 0.15 mJy beam$^{-1}$ for the 0\farcs06 and 0\farcs14 beams, respectively. While Ka-band was also
calibrated for polarimetry, we only provide results for Q-band since 
it has the brightest dust emission. We do not show the Stokes 
Q and U images for the sake of brevity, but describe their properties in Table \ref{image-parameters}.



\subsection{Millimeter to Radio Spectrum}

\begin{figure}
\begin{center}
\includegraphics[scale=0.5]{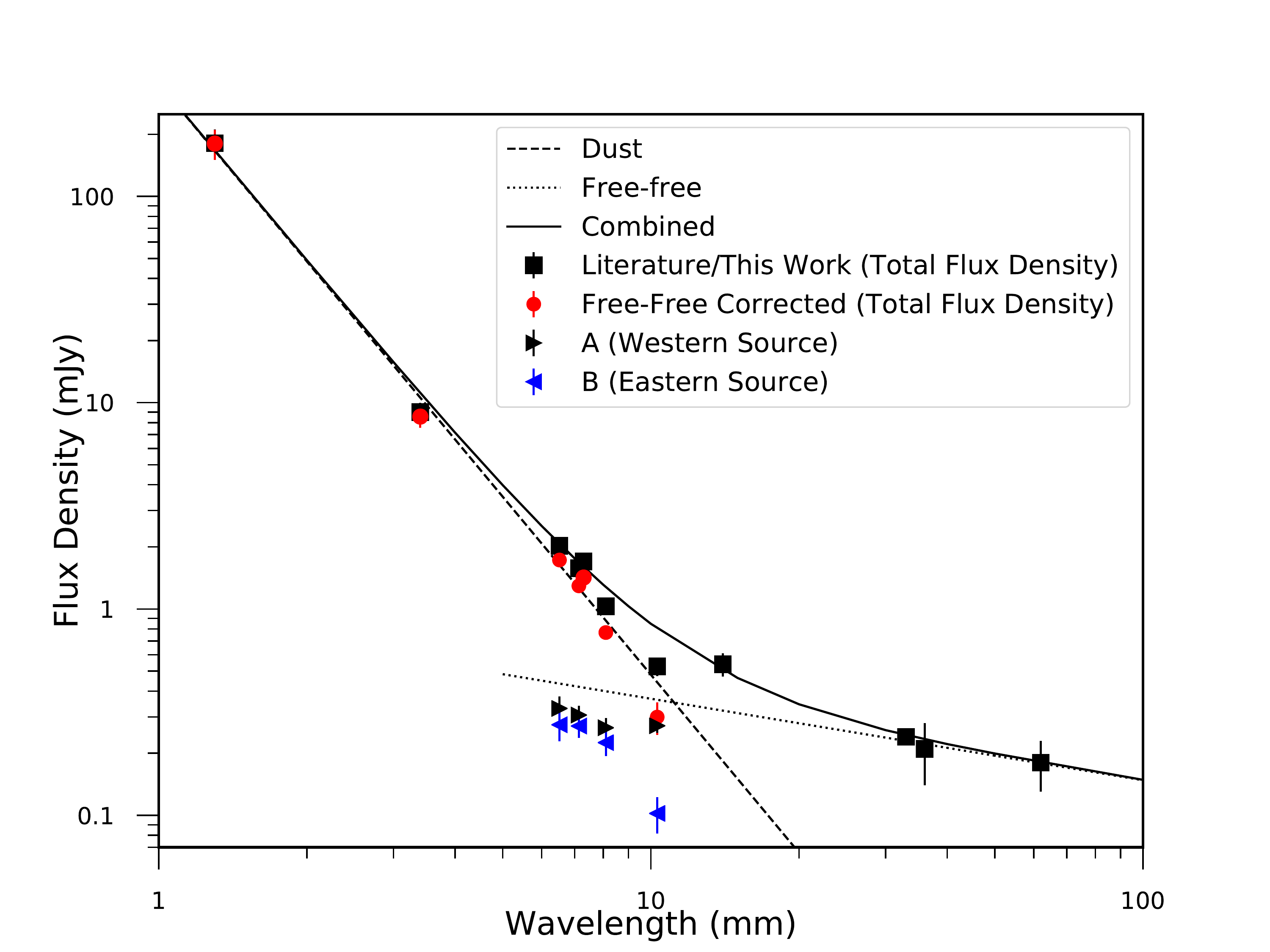}

\end{center}
\caption{Radio spectrum of L1157 MMS. The black points are the raw measurements of
the flux densities toward L1157 MMS as a whole, including both components. The spectrum
is modeled using two power-laws, one describing the free-free emission (dotted line) and
the dust emission (dashed line). The black line is the sum of the dust and free-free
fits. The red points show the emission with the estimated free-free emission subtracted.
Lastly, the individual flux densities for the A (black right pointing triangles) and 
B (blue left pointing triangles) components are plotted. 
The uncertainties shown on each flux density measurement
are statistical only.
}
\label{radio-spectrum}
\end{figure}

We have updated the radio
to millimeter spectrum of L1157 MMS with these new data, sampling 27 to 48~GHz. 
Furthermore, we also report the flux
densities of the two continuum peaks detected that likely trace a proto-binary
system. We measured the integrated flux densities in several wavelength bins,
but using lower angular resolution 
images with the beams approximately matched to the \textit{Naturally}-weighted beam at 44 to 48~GHz 
(see Table \ref{image-parameters}
and the Per 4 GHz - Binary Unresolved images for more detail).
We measured the flux
densities using Gaussian fitting with the CASA task \textit{imfit} to 
measure the integrated flux densities; the results from Gaussian fitting are consistent within
uncertainties to flux densities extracted from an aperture encompassing all emission.
The flux densities
are listed in Table \ref{flux-densities}, along with flux densities from the literature. The radio
spectrum shown in Figure \ref{radio-spectrum} has the typical form expected for a protostar, 
dust-dominated emission at the short wavelengths ($\lambda$~$<$1~cm) and free-free-dominated emission at 
longer wavelengths ($\lambda$~$>$1~cm).

We fit the radio spectrum with two power-laws, one corresponding
to the long-wavelength emission and another corresponding to the 
shorter wavelength emission. We first fit the free-free only portion 
of the spectrum at 3~cm to 6~cm since the
contribution of dust emission to those flux densities is negligible. The free-free
portion of the spectrum can be described by the function
$F_{\nu,ff}$~=~0.37($\nu/30~GHz)^{0.4\pm0.15}$. We then subtracted this
estimated free-free emission from all the data at $\lambda$~$<$~1~cm. Using the
free-free corrected data, we fitted the slope of the dust emission, finding
$F_{\nu,dust}$~=~0.48($\nu/30~GHz)^{2.87\pm0.1}$. Then we plot the 
sum of these two power-laws in Figure \ref{radio-spectrum}, and this function
matches the total flux densities of the source from 0.1~cm to 6~cm.

We also measured the flux densities toward the two resolved sources. We
used the \textit{imfit} task in CASA to fit two Gaussians simultaneously. We fixed
the source sizes to correspond to the synthesized beams in each image to limit the free
parameters in the fit and to avoid fitting the extended emission.
However, the residuals are negligible after subtracting the two point sources
indicating that these higher resolution images (Table 3; Per 4 GHz - Binary Resolved images)
are recovering less flux density than those with lower resolution (Table \ref{image-parameters}
and the Per 4 GHz - Binary Unresolved images). 
As such, the flux densities
of the two resolved components
are significantly lower than the integrated flux densities shown in Figure \ref{radio-spectrum}. 
Their spectral slopes are also more shallow, at least from 6.5~mm to 8.1 mm, which could
be due to the free-free emission contributing a more significantly fraction to the total emission for each component. Then, at
10.3~mm the flux density for B is low compared to the other measurements, while the flux density for A
is comparable to the 8.1~mm flux density. This result could be due to low S/N for B at 10.3~mm
where it appears less distinct than in the other bands. 

\subsection{Dust Mass}

The integrated flux density in the observed bands also enables
us to estimate the mass of the emitting material with the 
assumption of optically thin and isothermal dust emission using  
the equation
\begin{equation}
M_{dust} = \frac{F_{\nu} d^2}{\kappa_{\nu}B_{\nu}(T_{dust})}.
\end{equation}
$F_{\nu}$ is the integrated flux density at the chosen frequency, 
$d$ is the distance to the source, $\kappa_{\nu}$ is the dust mass
opacity at the same frequency as $F_{\nu}$, and $B_{\nu}(T_{dust})$
is the Planck function. We use the integrated flux density of 1.7 mJy
at 6.5~mm that corresponds to the flux density with 
the estimated free-free contribution removed (Table \ref{flux-densities}). We adopt a dust
temperature of $T_{dust}$=43~K (L/L$_{\rm bol}$)$^{0.25}$ \citep{tobin2020} which 
accounts for the luminosity dependence of the average dust temperature of the 
protostellar disk. L$_{\rm bol}$ is 7.6~\lsun\ \citep{green2013} when scaled to the revised distance of 352~pc, resulting in an estimated $T_{dust}$ $\sim$ 71~K, and we adopt $\kappa_{\nu}$=0.2~cm$^{2}$~g$^{-1}$ \citep{woitke2016} at 6.5~mm. With these assumptions,
we calculate that the dust mass of L1157 MMS is 372 \mearth\ of dust, and if
one adopts a dust to gas mass ratio of 1:100, the total mass is is calculated to
be 0.11~\msun. This calculated mass specifically applies to the size
scale of the emission detected here, all of which is contained within a radius of
50~au from the protostars.
Finally, comparable mass estimates can also be
derived from the 1.3~mm flux densities with the same assumptions, aside from the 1.3~mm
dust mass opacity.


\section{Discussion}

L1157 MMS is one of the most compact Class 0 multiple systems. Its separation of
$\sim$16~au is more compact than the detections found in
large surveys for protostellar multiplicity \citep{tobin2016,tobin2021}.
The wealth of complementary data on this system also enables us to put this detection
in the context of the known envelope properties \citep{looney2007,chiang2012, maury2019}, envelope rotation at small and larger scales \citep{tobin2011,gaudel2020}, the outflow \citep{bachiller2001,kwon2015,podio2016}, and magnetic fields \citep{stephens2013}.

\subsection{Inner Disk or Binary?}

An obvious question to ask is whether our detection of two peaks could simply be an inner disk or must it be a binary
protostar system. If the disk were a smooth circumstellar disk around a single source, it is expected to
appear centrally peaked. Other edge-on Class 0 disks observed with the VLA at 
comparable resolution appear centrally-peaked \citep[e.g.,][]{segura-cox2016,segura-cox2018,lin2021}.
However, if the disk had an inner cavity with a radius of approximately half the distance between
the components in L1157 MMS, it could present a similar appearance to a binary 
system with two disks. But then the cause of the central cavity would also 
require explanation, which would again point to a binary system with an even 
closer separation. Also, even in the case of a disk with a large central cavity,
we expect that the centroid of the free-free emission will be
in between the two dust peaks, which is not seen (see Appendix). 
Moreover, the flux density
of the western source does not decrease in the 35-39~GHz range, it is rather constant or slightly increased at 27-31~GHz (Figure \ref{radio-spectrum}). This is in line with free-free emission increasing at longer
wavelengths. To ultimately confirm that the free-free peaks have positions coincident the components
observed at Q and Ka-bands will require higher-resolution data at longer wavelengths.

Although we cannot fully rule-out a disk with a central inner cavity, we do not favor this interpretation.
Moreover, if our observations were tracing a single disk with such a large cavity, it would be difficult to
reconcile the large gap from the inner disk to the protostar with the previously observed large outflow rate of the protostar.  For that reason, we posit that L1157 MMS is a 16 au binary system, 
and we expect that the compact emission toward each component is tracing an unresolved circumstellar
disk in addition to some free-free emission.

\subsection{Formation Mechanism}
The presence of such a close binary in L1157 MMS raises immediate questions 
as to how it formed in this young system. L1157 MMS has a well-ordered magnetic field
exhibiting a clear hourglass shape as observed via polarimetry at 1.3~mm from 
the Combined Array for Millimeter Astronomy (CARMA) by \citet{stephens2013}.
The envelope specific angular momentum profile measured by
\citet{gaudel2020} clearly shows a decrease in specific angular momentum
from large to small radii, possibly indicating that magnetic braking has removed
angular momentum from the system. 
Thus, 
the binary system in L1157 MMS must have formed in the absence of both strong rotation
and a large disk.

One possibility is that the companion formed at a larger initial separation, 
and the accretion of low angular momentum material drove the separation 
smaller \citep{zhao2013}. The specific angular momentum of the binary system
is $\sim$8$\times$10$^{-5}$ km~s$^{-1}$~pc, assuming each component
is 0.02~\msun\ \citep{kwon2015}
estimated from their C$^{18}$O kinematic data that showed very
low rates of rotation on $<$1000~au scales. This value is consistent with the specific angular momentum profile
extrapolated to smaller radii by \citet{gaudel2020}. Thus, the scenario of forming
at larger radii and migrating is possible, and a companion could migrate from 100s
of au to 10s of au on time scales of a few 10s of kyr \citep{zhao2013}. However, 
the plausibility of this scenario is uncertain because the most practical method of 
fragmenting the envelope on these scales would be turbulent fragmentation \citep{offner2010,lee2019},
but the envelope itself has very narrow line widths \citep[$<$1~\kms,][]{tobin2011} and the line width increases in the inner envelope appear to be associated with the outflow rather than internal kinematics.
There is also no evidence of a second outflow having been launched within several hundred au of the main
protostars, which could have served as a signpost of past star formation. Instead, 
there is primarily evidence for twin jets with an overlapping origin point \citep[][see Section 4.3]{kwon2015}.

The other possibility is that the companion formed near its current location within
a small disk. If we assume that the total flux density at 6.8~mm corresponds to the
emission from the original single disk, then the disk mass was
$\sim$0.11~\msun.
We can estimate the stability of that disk around L1157 MMS using a simplified
form of Toomre's Q parameter from
\citet{kratter2016}
\begin{equation}
\label{eq:qapprox}
Q \approx 2\frac{M_*}{M_d}\frac{H}{R}.
\end{equation}
$M_{*}$ is the mass of the protostar (in solar masses), $M_d$ is the 
total mass of the disk (gas and dust) in solar masses, H is the vertical
scale height of the disk, and R is the radius of the disk. The vertical scale
height $H$ is equivalent of $c_s$/$\omega$; $c_s$ is the sound speed, and
$\omega$ is the angular velocity at the radius where Q is being measured.
We use an estimated rotation rate of the two protostars of 1.0~\kms\ at 16~au, translating
to $\omega$ = 4$\times$10$^{-10}$~s$^{-1}$, and we use $c_s\sim$0.54~\kms\ at a temperature
of $\sim$71~K. Then the total $M_{*}$ is estimated to be 0.04~\msun\ \citep{kwon2015}.
Plugging
these terms into the equation yields Q$\sim$0.4, indicating that the original inner disk
around L1157 MMS could have been gravitationally unstable. However, uncertainties 
in translating the 6.8~mm flux density to total disk mass, the protostar mass,
and the disk temperature make this a very tentative estimate, but this calculation does demonstrate
that the inner disk perhaps had enough mass to fragment.
The disk in the past could have had even higher mass prior
to fragmentation, much of which could have been accreted to form the a secondary protostar.

The formation of the companion \textit{in situ} would require that a compact, gravitationally
unstable disk formed within L1157 MMS. Moreover, if the observational constraints on the 
magnetic field and rotation rates are correct, the rotation of the envelope could have 
undergone significant magnetic braking, restricting disk formation to a relatively small
radius where non-ideal MHD effects like Ohmic dissipation and the Hall effect could dissipate
the magnetic flux and enable disk formation \citep[e.g.,][]{li2014}. Simulations of disk formation within
magnetically-dominated envelopes do show that formation of gravitationally unstable
disks with radii of only 10s of au are indeed possible \citep{zhao2018,lam2019, xu2021}. Therefore,
the idea that the companion formed \textit{in situ} via fragmentation of a compact, gravitationally unstable
disk is plausible.


    

\subsection{Binary and Outflow}

One of the most obvious features of the large-scale outflow from 
L1157 MMS is that it appears to be precessing. This was apparent in the
earlier data from \citet{gueth1996,bachiller2001} and in 
the \textit{Spitzer Space Telescope} images of the shock-excited molecular hydrogen emission with a
ribbon-like appearance \citep{looney2007}. More recently, \citet{kwon2015} 
modeled the outflow as two precessing jets, and \citet{podio2016} modeled the outflow
as a single precessing jet. 

The two jet fit from \citet{kwon2015} implicitly assumed that there
is a binary system launching the jet and that the binary system is causing
tidal precession. From the modeling, they
found that for the two jet fit the orbital period is 
(50–370) M$_1^{-1/4}$yr, while for Jet 2 the orbital period is (60–450) 
M$_1^{-1/4}$yr. The orbital radius for Jet 1 is (13–52) M$_1^{1/3}$~au
and for Jet 2 the orbital radius is (15–60) M$_1^{1/3}$~au. The mass
of each component was assumed to be 0.02~\msun, totaling 0.04~\msun\ 
(see section 4.2).
The lower range of orbital
radii fit are consistent with the observed separation of the two continuum sources.
Even if the overall mass of the protostars is higher, 
the orbital radii of the jets only increase slowly with increasing mass.

The orbital motion
of the binary alone does not cause the outflow precession as the orbital 
periods are much shorter than the apparent outflow precession. Thus, \citet{kwon2015} found that the
precession periods are $\sim$5200 and $\sim$7600 yr for the two jets. On
the other hand, \citet{podio2016} finds an outflow precession period
of 1640~yr. The assumption of a single outflow in \citet{podio2016} 
makes it necessary for the outflow to precess faster in order to match the
observations.

The orbital radii and periods of the jets are compatible with the projected separation
of the two continuum sources that we detect. Moreover, the projected separation
could be smaller than the full semi-major axis since we do not know where
in the orbital path we are observing the system. Therefore, the observed parameters of the binary
system are completely consistent with the constraints from modeling of the outflows.

\subsection{Context with other Binary systems}

L1157 MMS is 
the most compact known Class 0 proto-multiple system
resolved in dust continuum emission. Other 
similarly compact proto-multiple systems are IRAS 03292-3039 (Per-emb 2) (24~au),
HOPS-361-E (22~au),  BHB2007 11 (28~au), Per-emb 18 (26~au), VLA1623A (30~au), and L483 (30~au) \citep[][Cox et al. subm.]{,tobin2016,harris2018,alves2019,tobin2021}. 
The protostar IRAS 16293-2422A has also had a number of components
and potential substructures resolved in submillimeter to centimeter radio
continuum \citep[e.g.,][and references therein]{hernandezgomez2019,maureira2020,oya2021} that can have separations as close as $\sim$0\farcs1. However, the most robust
components that are likely to be true protostellar sources are the A1 and 
A2 sources that are $\sim$54~au in projected separation.

The primary difference between L1157 MMS and these other systems, other than its
very close separation, is the presence of obvious rotation
toward most of them (except HOPS-361-E), likely corresponding to circumbinary disk emission
or at least a rotating envelope \citep{tobin2018,alves2019}. L1157 MMS on the other hand shows no signs of strong rotation on $<$~1000~au scales as discussed in Sections 1 and 4.2. Thus, 
L1157 MMS demonstrates that fragmentation and close multiple formation
is still possible in cases where the dynamics of the collapse may be
regulated by magnetic fields. 

We emphasize that the case of L1157 MMS is distinct from NGC 1333 IRAS 4A (IRAS4A), which is another proto-binary with 
a well-ordered magnetic field \citep{girart2006,cox2015,ko2020}. The binary in IRAS4A has a much wider separation
 of $\sim$540~au vs. 16~au for L1157 MMS. Furthermore, the magnetic field in IRAS4A is misaligned with respect
 to the outflows from the IRAS4A sources \citep{hull2014,chuang2021}, where as the magnetic field in L1157 MMS
 is aligned with the outflow.

The presence of such a close binary system in L1157 MMS demonstrates a pathway for
the formation of binary systems with separations $<$10~au. 
Currently the only evidence for protostellar multiple systems with separations
$<$10~au are a few radial velocity observations of Class I protostars
\citep{vianaalmeida2012} and the indirect evidence from variability 
in one system that is interpreted as binary-modulated pulsed accretion \citep{muzerolle2013}.
More-evolved Class II and Class III pre-main-sequence systems, on the other hand,
have multiplicity characteristics at $<$10~au consistent with field stars \citep{kounkel2019}.
Thus, it is an open question whether these $<$10~au companions
first formed at much larger separations (many 10s to 100s of au) and migrated inward 
or if they could have formed closer in. The 16~au binary in L1157 at least suggests that
formation near 10~au separation is possible in Class 0 systems and migration
of clumps and/or fully formed companions could easily populate $<$10~au separations \citep[e.g.,][]{zhu2012,zhao2013}.


\section{Conclusions}
We have discovered a compact proto-binary system forming within L1157 MMS using
observations at Q (6.8~mm) and Ka-bands (9~mm) with the VLA. It is 
currently the most compact known Class 0
proto-binary system, with a projected separation of $\sim$16~au. The mass
of the inner envelope and disk surrounding the protostars (within a 50~au radius)
is estimated to be $\sim$0.11~\msun, using simple assumptions 
for converting dust emission to mass. The 
L1157 MMS system as a whole has a large flattened envelope with very weak rotation
from 10000~au to $\sim$200~au scales probed by both \nthp\ and \cateo\ emission \citep{tobin2011, kwon2015,
gaudel2020}, and the system has an observed poloidal magnetic field \citep{stephens2013}. Thus, L1157 MMS also
demonstrates that the formation of compact binary systems is possible, even when
magnetic fields may be dynamically important to the collapse and disk formation.
Finally, of the known multiple protostar systems with separations $<$~30~au that have
had their inner envelope/disk kinematics characterized, L1157 MMS stands alone in being
the only one without clear organized rotation surrounding the binary.

\acknowledgements
We thank the anonymous referee for a constructive report that
improved the quality of the paper.
We acknowledge useful discussions with Z.~Li regarding the
interpretation of the system. J.J.T. acknowledges support from  NSF AST-1814762.
L.W.L. acknowledges support from NSF AST-1910364 and AST-2108794.
The National Radio Astronomy Observatory is a facility of the National Science
Foundation operated under cooperative agreement by Associated Universities, Inc.

 \facility{VLA}
\software{Astropy \citep[http://www.astropy.org; ][]{astropy2013,astropy2018}, 
APLpy \citep[http://aplpy.github.com; ][]{aplpy}, CASA \citep[http://casa.nrao.edu; ][]{mcmullin2007}}

\appendix
\section{Robustness of the Binary}

\begin{figure}
\begin{center}
\includegraphics[scale=0.475]{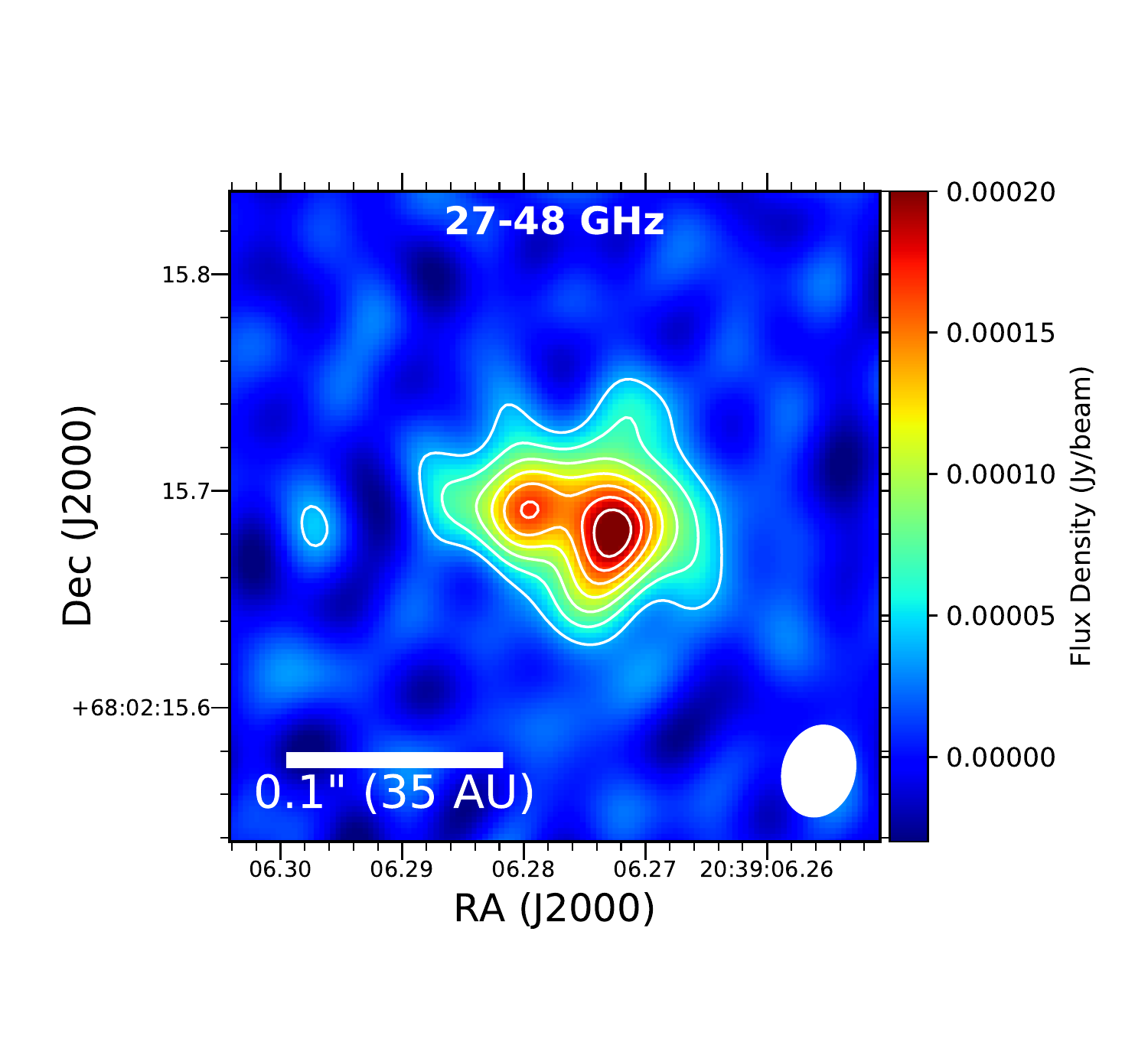}
\includegraphics[scale=0.475]{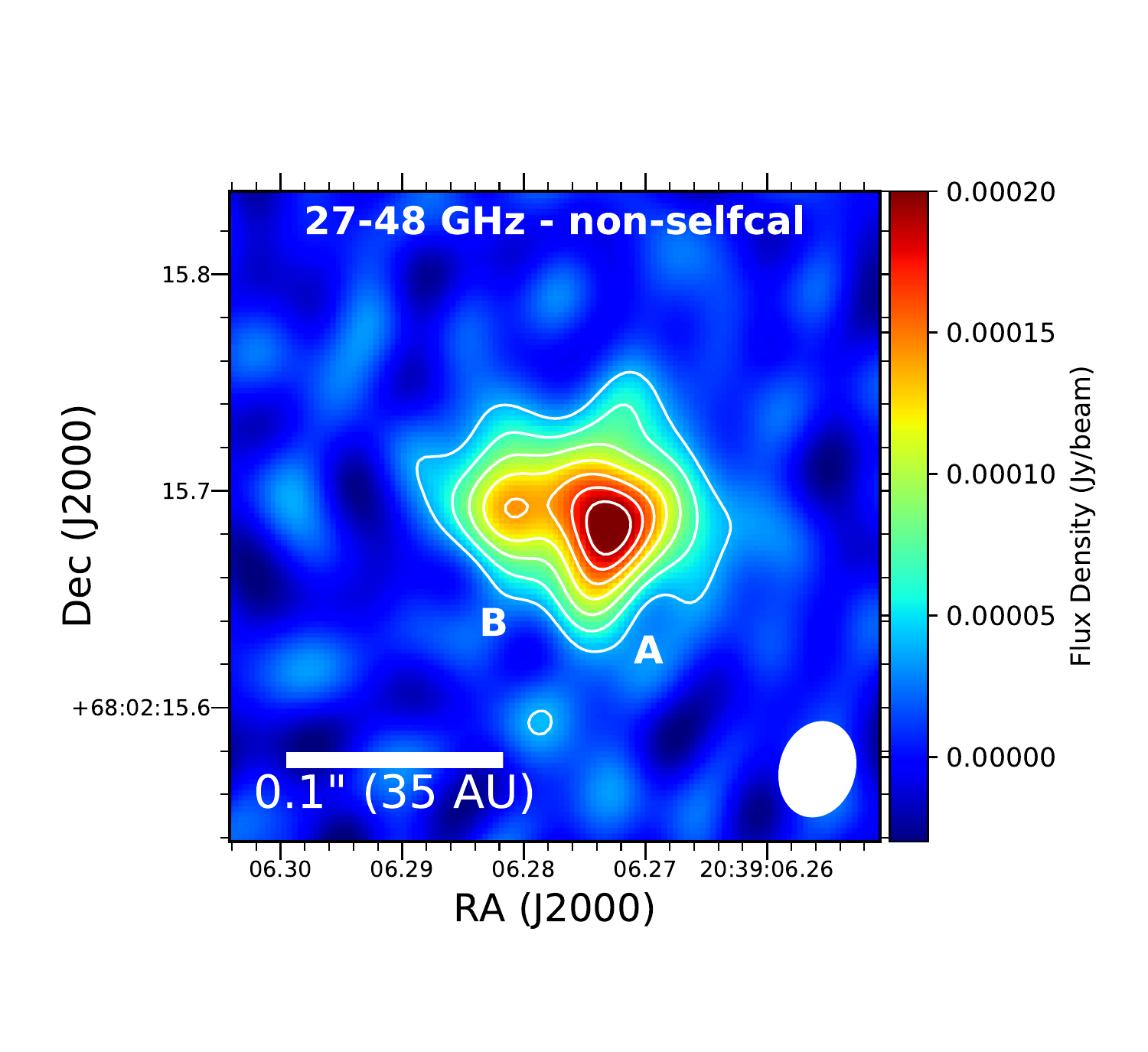}
\includegraphics[scale=0.475]{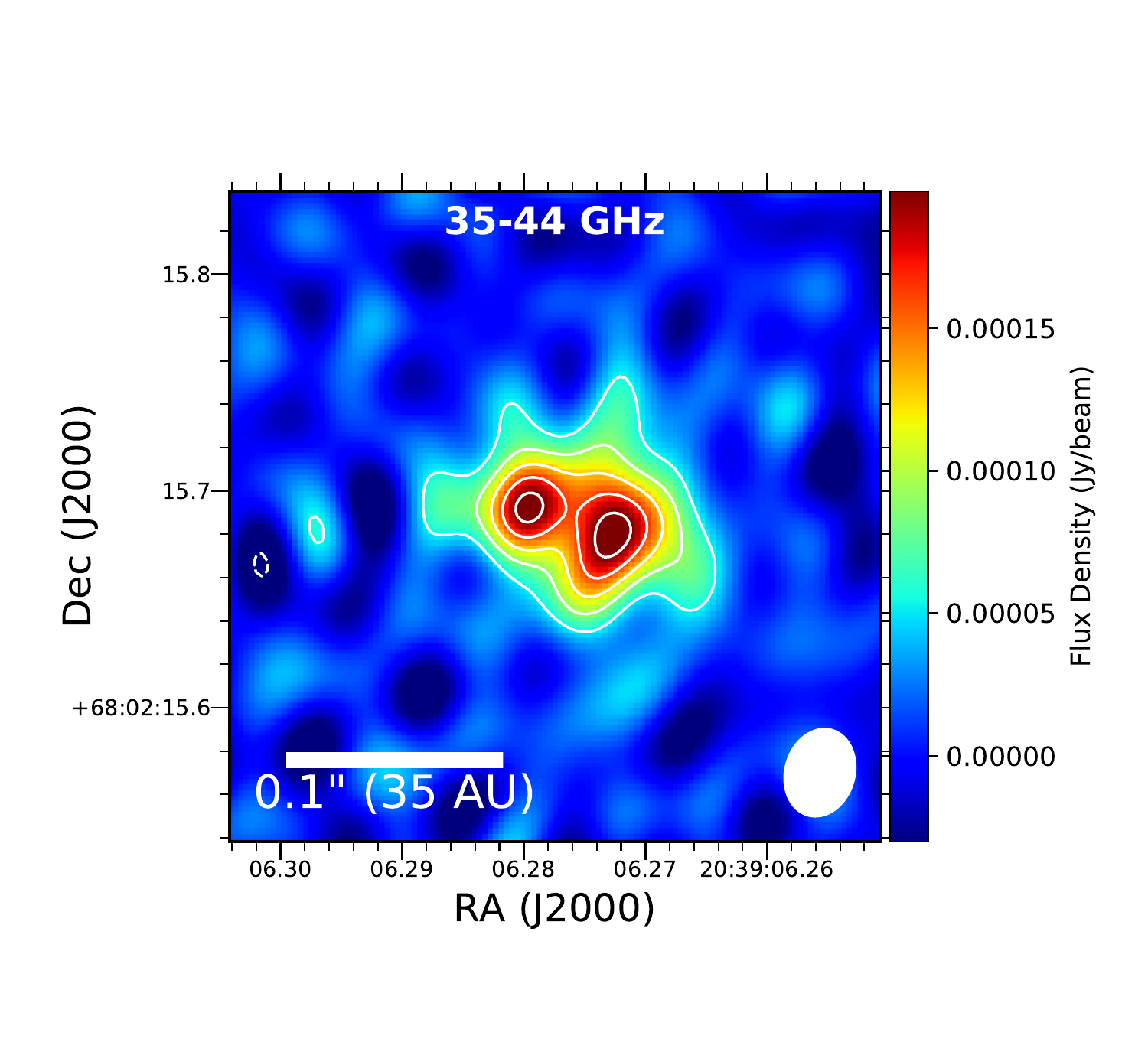}
\includegraphics[scale=0.475]{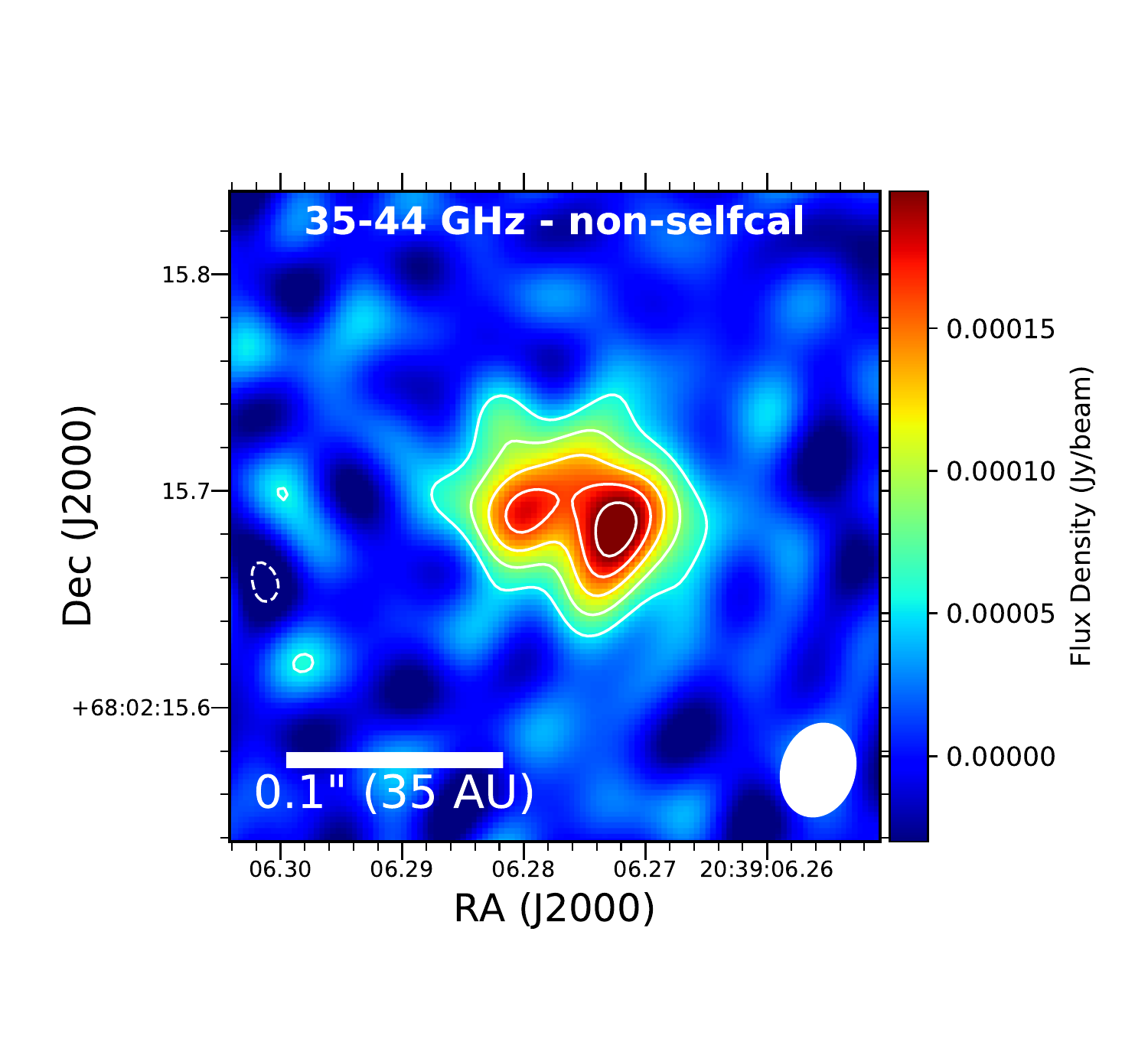}

\end{center}
\caption{Images of L1157 MMS generated using the full combined Q and Ka-bands (top) and 
only the 35 to 44~GHz spectral range (bottom) with a robust parameter of -0.25. We also
show the self-calibrated images (left) in comparison to the images \textit{without} self-calibration (right).
The binary clearly stands out in the self-calibrated images better than the non-self-calibrated images,
with its significance increasing by $>$2$\sigma$. The contours
in the right panels start at 3$\sigma$ and increase on 2$\sigma$ intervals;
values for $\sigma$ and the beam sizes in each image are provided in Table \ref{image-parameters}.
The noise and beam sizes are approximately the same for the self-calibrated and non-self-calibrated images.
}
\label{continuum-highres}
\end{figure}

\begin{figure}
\begin{center}
\includegraphics[scale=0.475]{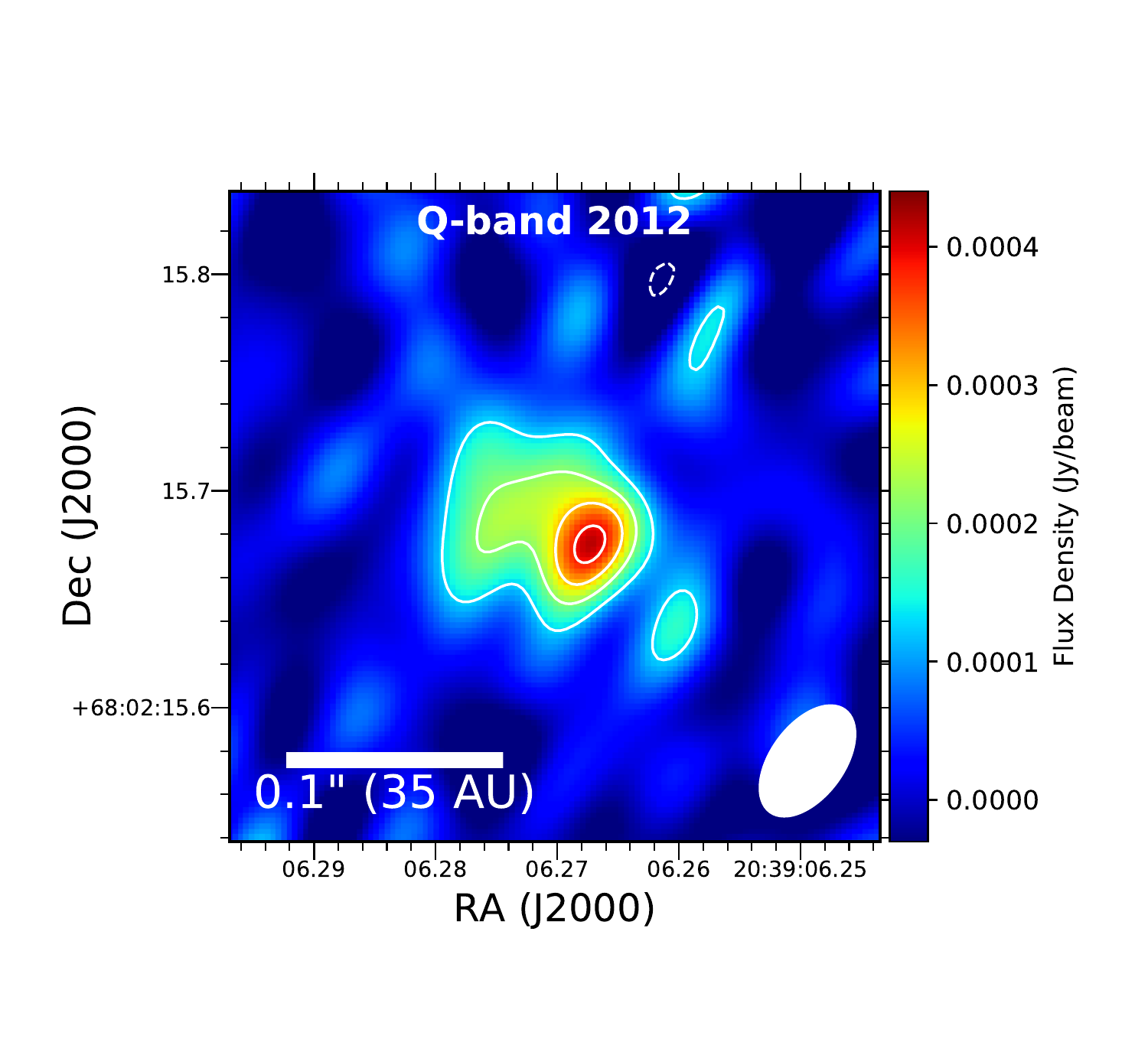}
\end{center}
\caption{Images of L1157 MMS at 41~GHz from the observations in 2012.
This is re-reduced from the data presented in \citet{tobin2013}. This
image was produced with robust=-0.5. The contours are drawn at -3, 3, 5, 7, 9$\sigma$,
where $\sigma$=43~$\mu$Jy~beam$^{-1}$. The beam is 0\farcs058$\times$0\farcs033. The 
morphology is consistent with the newer data, lending confidence that the
second continuum peak is a robust feature.
}
\label{continuum-2012}
\end{figure}

We performed several additional tests on
the data to confirm the robustness of the binary detection.
We first re-reduced the data from \citep{tobin2013} and show the data in
Figure \ref{continuum-2012}. These data were Q-band, had
2~GHz of bandwidth, and only include the A-configuration data from
the paper. The image clearly shows that L1157 MMS is extended, in the same
direction as detected in Figures \ref{continuum-highres}, showing evidence for
two peaks. However, the eastern source is not a distinct point source owing 
to the lower sensitivity of those data, but like the data in Figure \ref{continuum-highres}
the western source is brighter.

\begin{figure}
\begin{center}
\includegraphics[scale=0.475]{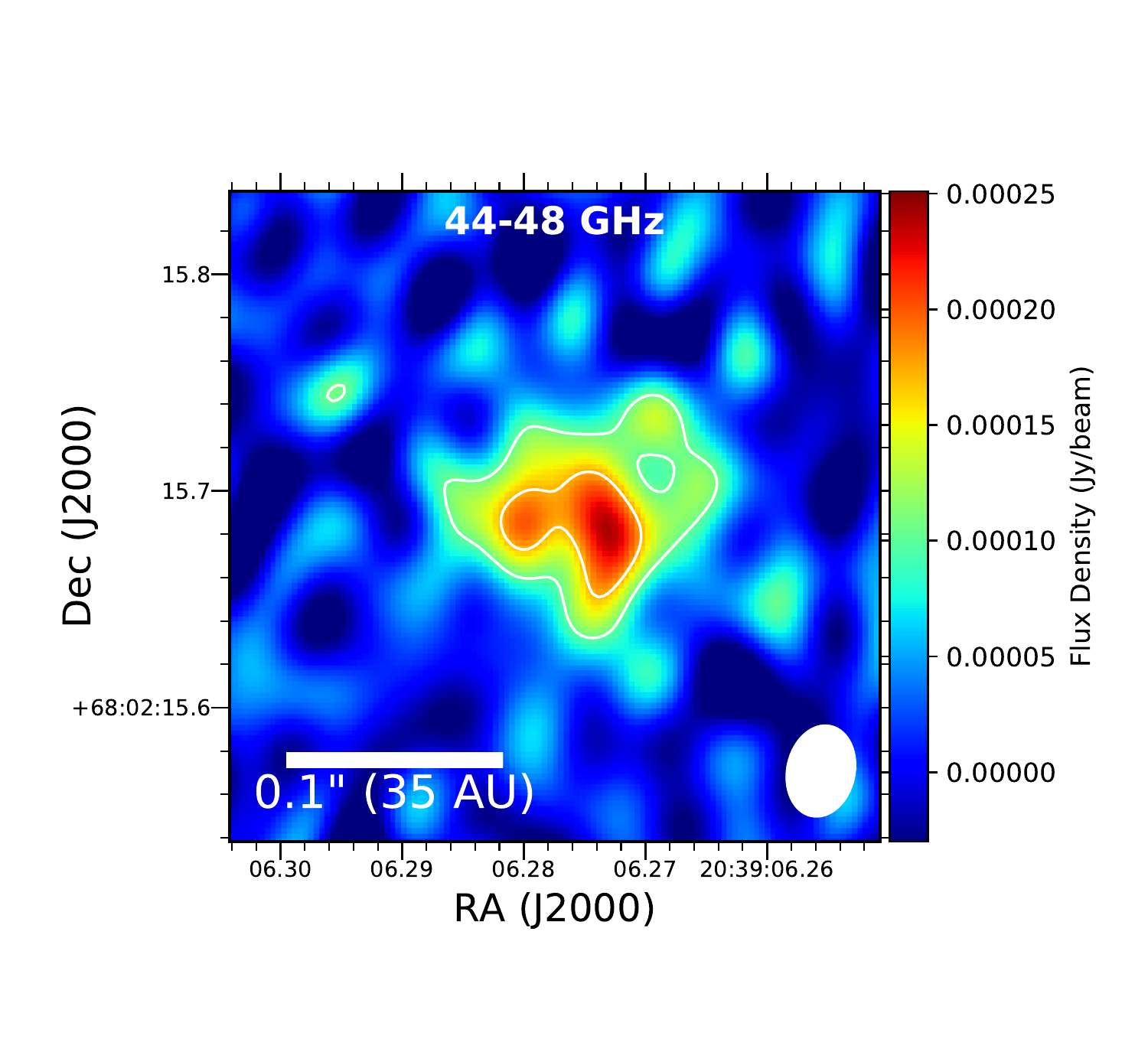}
\includegraphics[scale=0.475]{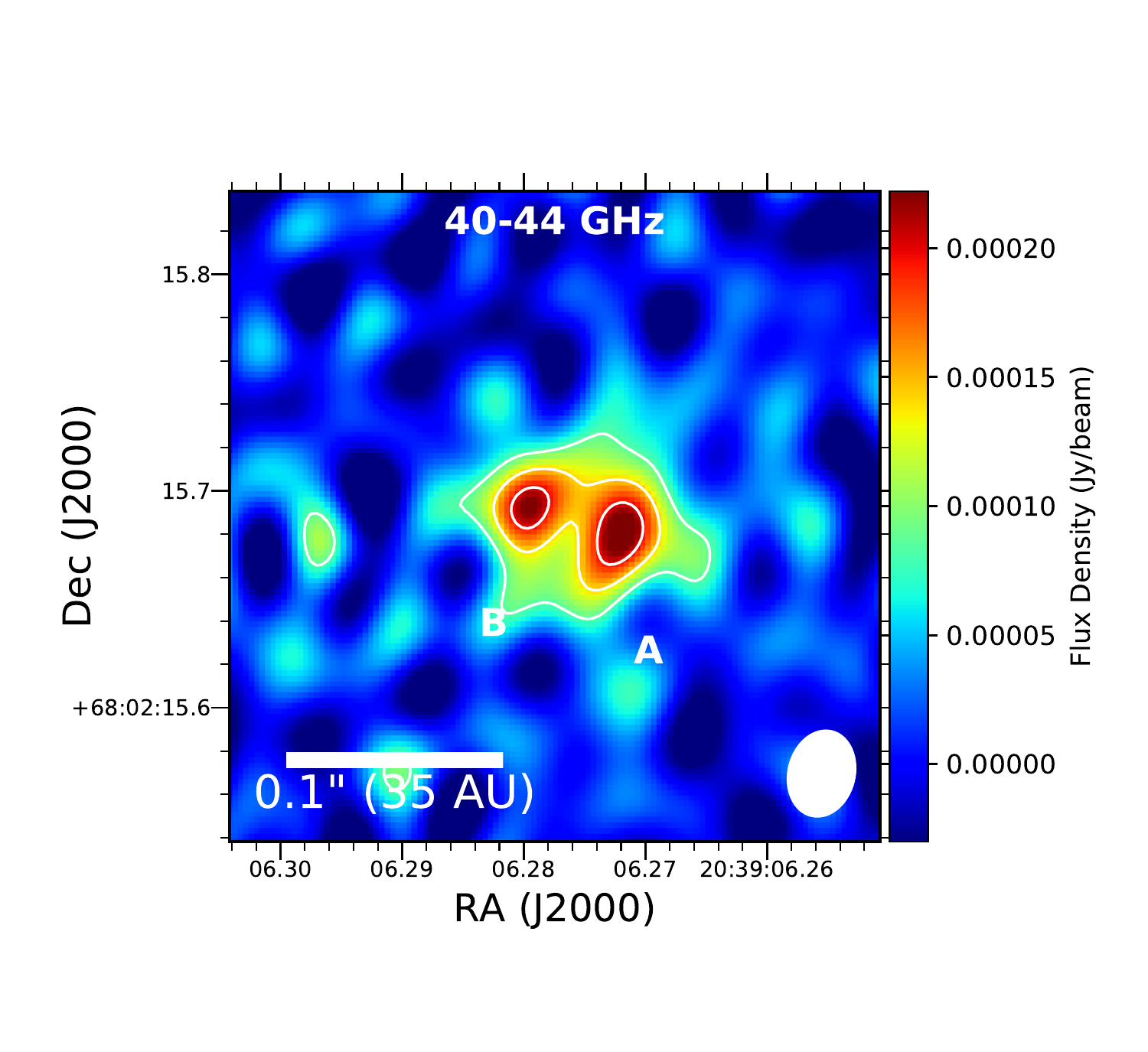}
\includegraphics[scale=0.475]{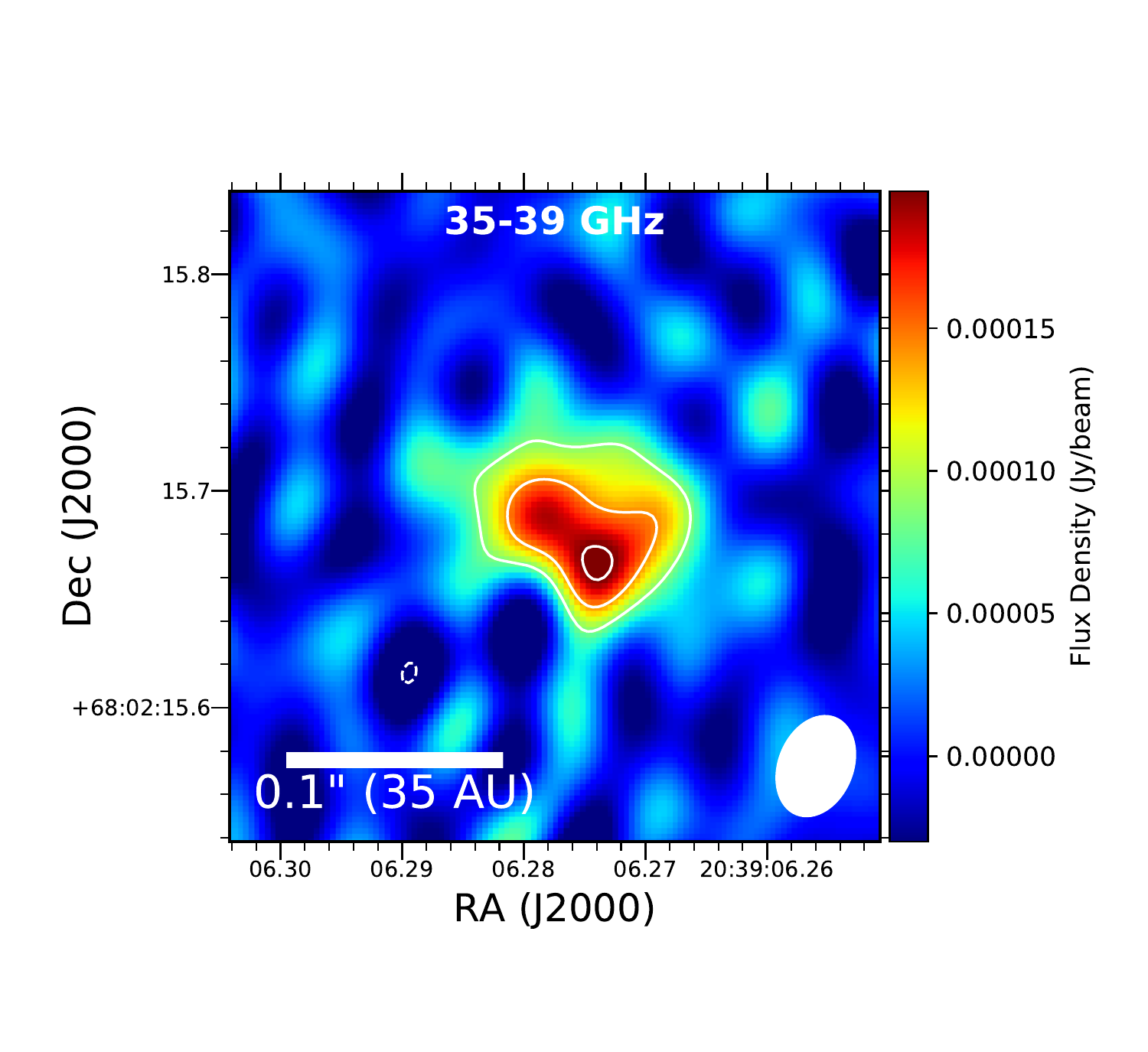}
\includegraphics[scale=0.475]{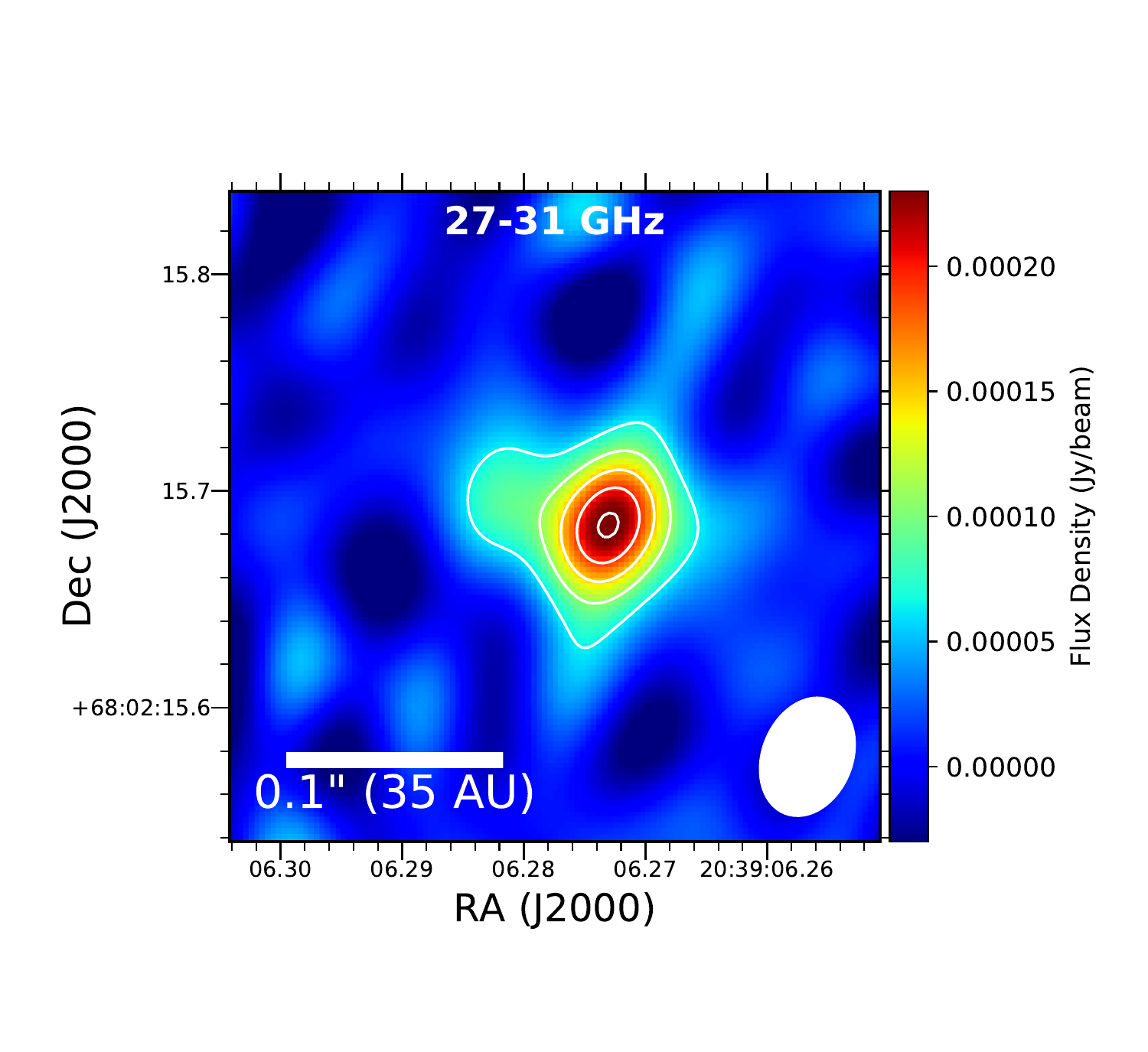}
\end{center}
\caption{Images of L1157 MMS in 4 GHz chunks across the observed bands. The contours start at $\pm$3$\sigma$ and
increase on 2$\sigma$ intervals, where $\sigma$ and the beams for each image are provided in Table \ref{image-parameters}.
}
\label{continuum-multiwave}
\end{figure}

We then further analyzed the data presented in this paper to examine the robustness
of the observed binary. To ensure that the detection is robust across our 
observed frequency range, we separately imaged
4~GHz chunks of the observed bands from the combined dataset. At each wavelength
we tuned the $robust$ parameter to provide similar beams in each frequency interval (Table \ref{image-parameters}).
We show the images
in Figure \ref{continuum-multiwave}, and each frequency interval
similarly shows a double peaked structure. The lowest frequency band
only marginally resolves the structure because of its inherently lower angular
resolution. Moreover, the peak intensity of the western peak is $\sim$1.5 to 2$\times$
brighter than the eastern peak. 

\begin{figure}
\begin{center}
\includegraphics[scale=0.475]{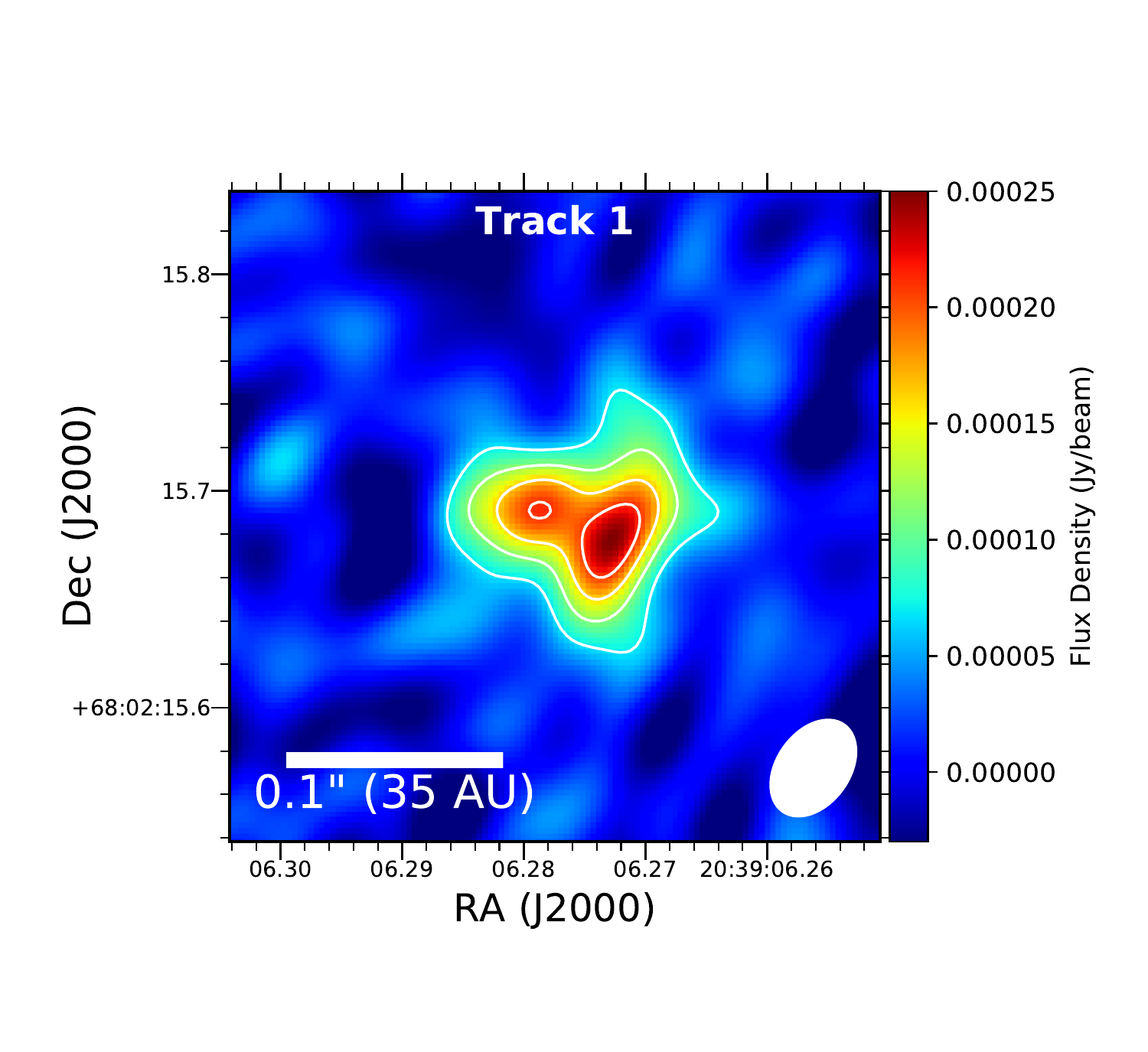}
\includegraphics[scale=0.475]{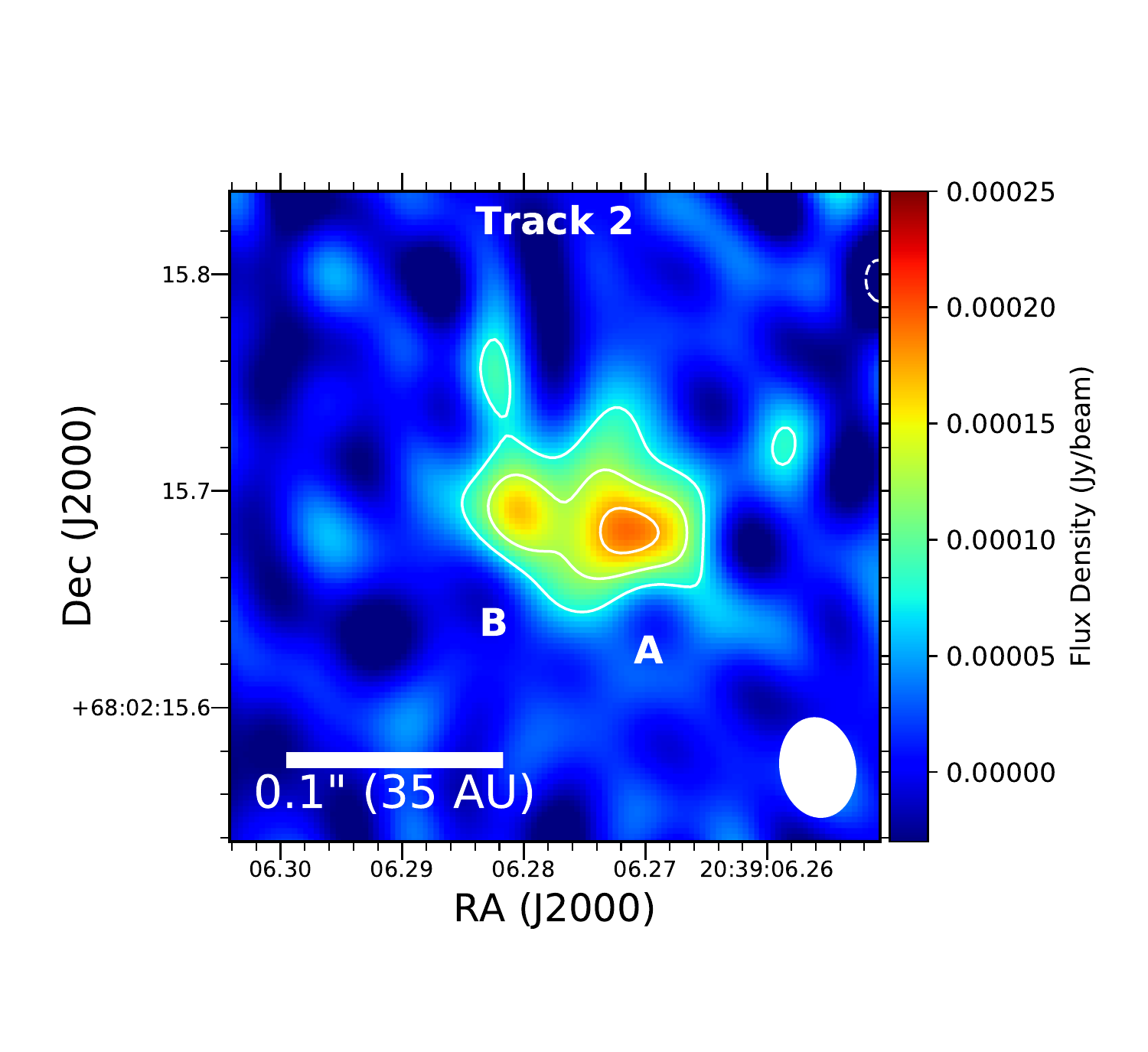}
\includegraphics[scale=0.475]{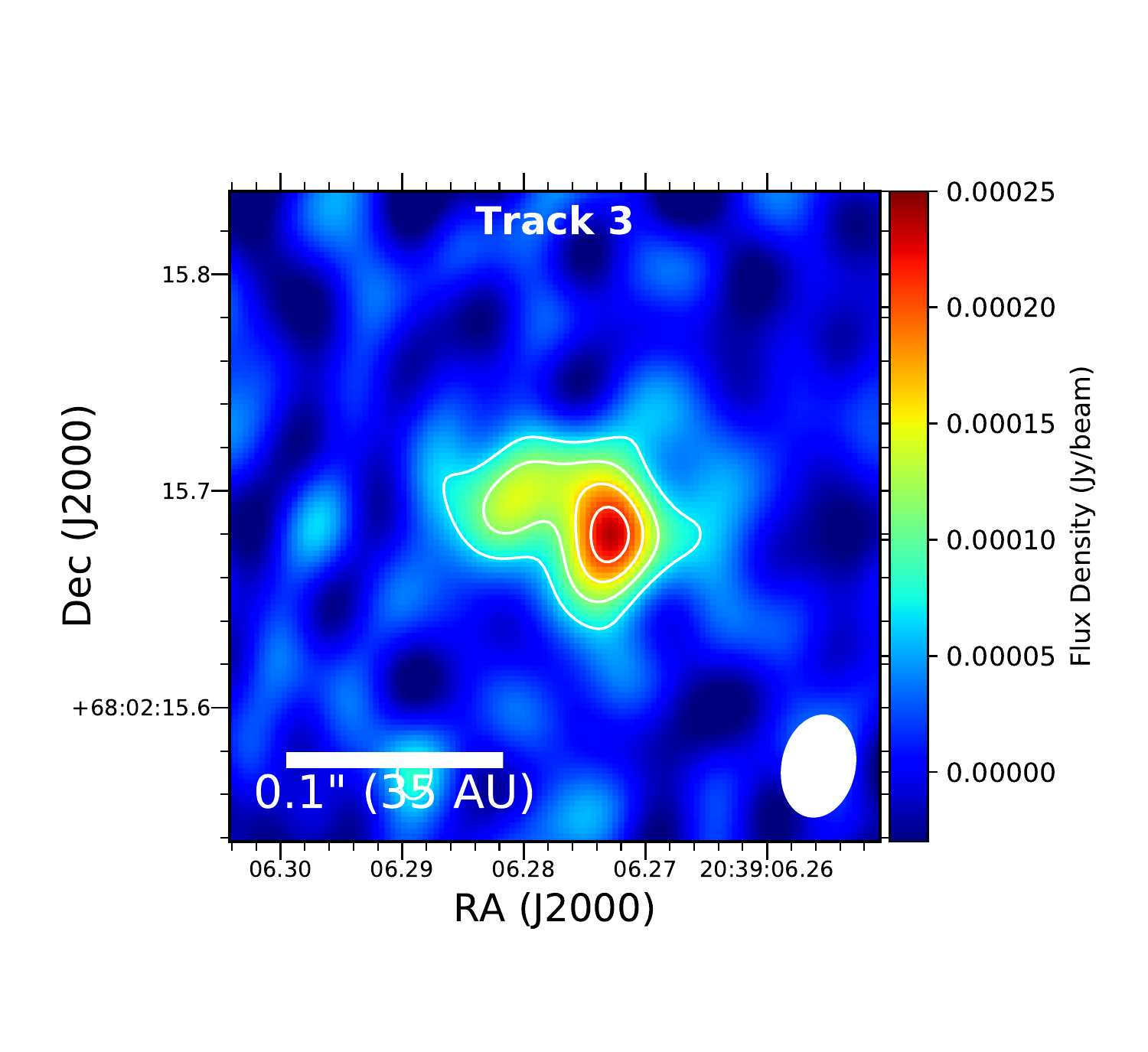}
\includegraphics[scale=0.475]{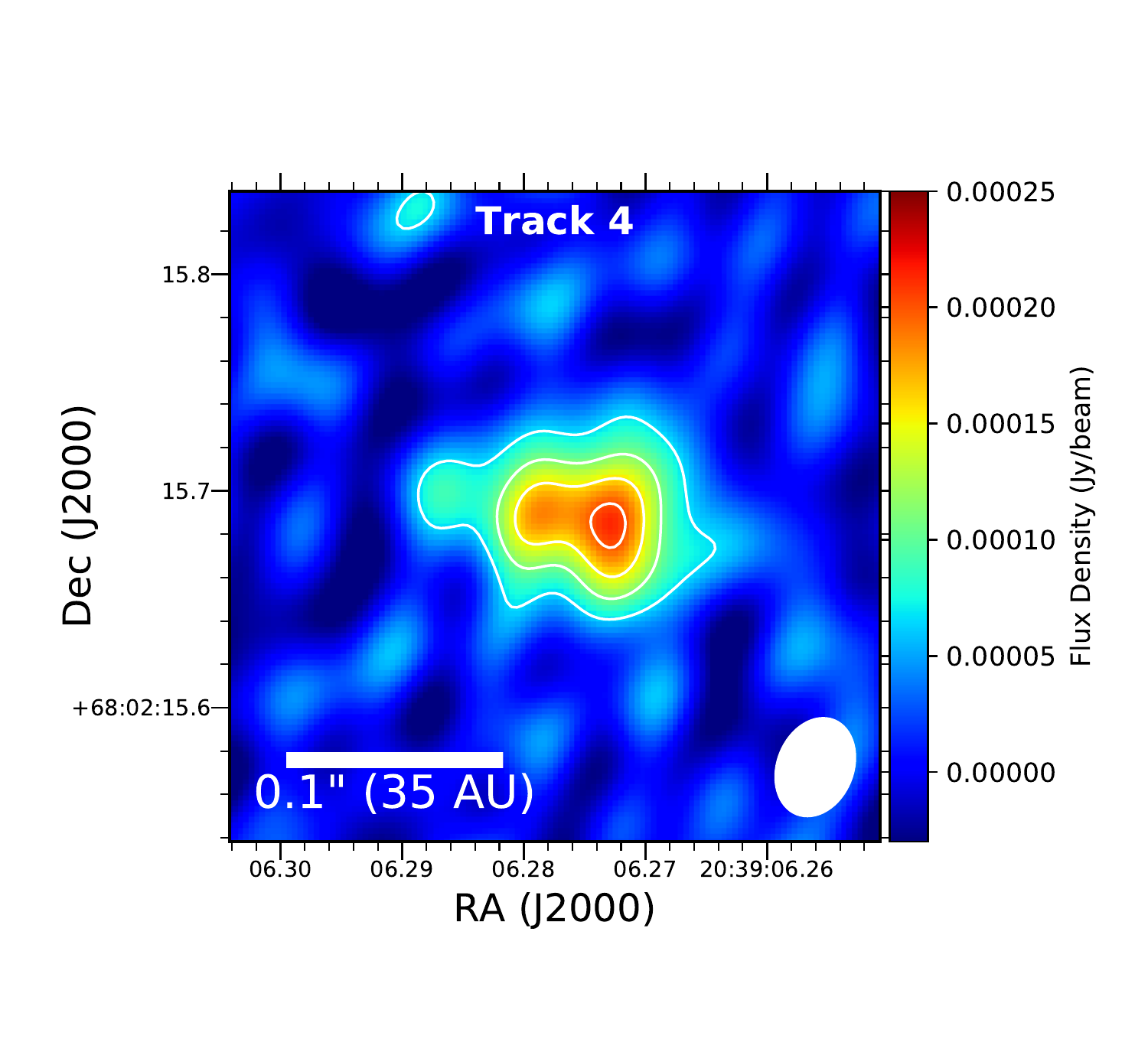}
\end{center}
\caption{Images of L1157 MMS from each of the 4 epochs of observation using the combined Q and Ka-band spectral range.
The contours start at $\pm$3$\sigma$ and
increase on 2$\sigma$ intervals, where $\sigma$ and the beams for each image are provided in Table \ref{image-parameters}.
}
\label{continuum-multiepoch}
\end{figure}

We next imaged each observation from Table 1 separately, using the full Q and Ka-band bandwidth
(Figure \ref{continuum-multiepoch}).
While there is some variation in the observed structure in each epoch, which 
is expected from phase noise, there is a double-peaked structure detected
in all epochs of observation. Thus, the two peaks resolved in the new data, 
in conjunction with a similar structure resolved in the previous
data, and two peaks resolved with different VLA bands,
lead us to conclude that this resolved structure is robust
and is not an artifact of low S/N or phase noise in the data.

\begin{small}
\bibliographystyle{apj}
\bibliography{ms}
\end{small}

\clearpage

\clearpage

\begin{deluxetable}{llccll}
\tabletypesize{\scriptsize}
\tablewidth{0pt}
\tablecaption{VLA Observation Log}

\tablehead{\colhead{Fields} & \colhead{Date} & \colhead{Duration} & \colhead{Calibrators}\\
                            &                & \colhead{(hr)}     &\colhead{Bandpass, Flux, Complex Gain}
}
\startdata
L1157   & 2015 Jun 26 & 2.5 &  3C84, 3C48, J2006+6424\\
L1157   & 2015 Jun 30 & 2.5 &  3C84, 3C48, J2006+6424\\
L1157   & 2015 Jul 01 & 2.5 &  3C84, 3C48, J2006+6424\\
L1157   & 2015 Sep 21 & 2.5 &  3C84, 3C48, J2006+6424\\
\enddata
\tablecomments{}
\label{vla_obs}

\end{deluxetable}

\begin{deluxetable}{lll}
\tabletypesize{\scriptsize}
\tablewidth{0pt}
\tablecaption{Source Positions}

\tablehead{\colhead{Source} & \colhead{Right Ascension} & \colhead{Declination}\\
                            & \colhead{(J2000)} & \colhead{(J2000)}
}
\startdata
2015 Data\\
L1157 MMS & 20:39:06.2746$\pm$0.0001& +68:02:15.6853$\pm$0.001 \\
\hline
L1157 MMS-A & 20:39:06.2724$\pm$0.0001 & +68:02:15.680$\pm$0.002 \\
L1157 MMS-B & 20:39:06.2798$\pm$0.0001 &  +68:02:15.693$\pm$0.002\\
\hline
\hline
2012 Data\\
L1157 MMS & 20:39:06.2704$\pm$0.0001 & +68:02:15.683$\pm$0.002 \\
\hline
L1157 MMS-A & 20:39:06.2675$\pm$0.0001 & +68:02:15.674$\pm$0.002 \\
L1157 MMS-B & 20:39:06.2747$\pm$0.0002 &  +68:02:15.693$\pm$0.003\\
\enddata
\tablecomments{}
\label{sources}

\end{deluxetable}

\begin{deluxetable}{lllllllll}
\tabletypesize{\scriptsize}
\tablewidth{0pt}
\tablecaption{Image Properties}

\tablehead{\colhead{Image} & \colhead{Track} & \colhead{Stokes} & \colhead{Robust, Taper} & \colhead{Beam Size}& \colhead{RMS} & \colhead{Freq. Range} & \colhead{$\lambda$}& \colhead{Figure}\\
                            &  &   & & \colhead{(\arcsec)} & \colhead{($\mu$Jy~beam$^{-1}$)} &\colhead{(GHz)}& \colhead{(mm)} & 
}
\startdata
2015 Data\\
\hline
1.  Q+Ka & All Tracks  & I & 2.0 & 0.074$\times$0.060 & 6.9   & 27 - 48 & 8.45 & \ref{continuum-lowres} \\
2.  Q+Ka & All Tracks & I & -0.25 & 0.044$\times$0.034 & 12.6 & 27 - 48 & 8.45 & \ref{continuum-highres} \\
3.  Q+Ka & All Tracks  & I & 2.0 & 0.068$\times$0.057 & 9.8 & 35 - 44 & 7.6 & \ref{continuum-lowres}\\
4.  Q+Ka & All Tracks & I & -0.25 & 0.043$\times$0.033 & 17.8 & 35 - 44 & 7.6 & \ref{continuum-highres}  \\
5.  Q   & All Tracks & I & 2.0 &    0.061$\times$0.050 & 24.2   & 40 - 48 & 6.8 & \nodata\\
6.  Q   & All Tracks & Q & 2.0 &    0.061$\times$0.050 & 24.2   & 40 - 48 & 6.8 & \nodata\\
7.  Q   & All Tracks & U & 2.0 &    0.061$\times$0.050 & 23.2   & 40 - 48 & 6.8 & \nodata\\
8.  Q   & All Tracks & I & 2.0, 1000~k$\lambda$ &    0.14$\times$0.12 & 38.8   & 40 - 48 & 6.8 & \nodata\\
9.  Q   & All Tracks & Q & 2.0, 1000~k$\lambda$ &    0.14$\times$0.12 & 35.2   & 40 - 48 & 6.8 & \nodata\\
10. Q   & All Tracks & U & 2.0, 1000~k$\lambda$ &    0.14$\times$0.12 & 31.8   & 40 - 48 & 6.8 & \nodata\\
\hline
Per 4 GHz  - Binary Resolved\\
\hline
11. Q  & All Tracks & I & 0.0 & 0.043$\times$0.032 & 34.5 & 44 - 48 & 6.52 & \ref{continuum-multiwave}\\
12. Q  & All Tracks & I & 0.0 & 0.041$\times$0.031 & 27.8 & 40 - 44 & 7.14 & \ref{continuum-multiwave}\\
13. Ka & All Tracks & I & -0.5 & 0.049$\times$0.035 & 28.7 & 35 - 39 & 8.1 & \ref{continuum-multiwave}\\
14. Ka & All Tracks & I & -0.5 & 0.058$\times$0.043 & 21.5 & 27 - 31 & 10.3 & \ref{continuum-multiwave}\\
\hline
Per 4 GHz - Binary Unresolved\\
\hline
15. Q  & All Tracks & I  & 2.0 & 0.062$\times$0.049 & 23.6 & 44 - 48 & 6.52 & \nodata\tablenotemark{a}\\
16. Q  & All Tracks & I  & 1.5 & 0.060$\times$0.051 & 15.8 & 40 - 44 & 7.14 & \nodata\tablenotemark{a}\\
17. Ka & All Tracks & I & 0.5 & 0.062$\times$0.049 & 14.3 & 35 - 39 & 8.1 & \nodata\tablenotemark{a}\\
18. Ka & All Tracks & I & 0.0 & 0.063$\times$0.046 & 14.9 & 27 - 31 & 10.3 & \nodata\tablenotemark{a}\\
\hline
Per Observation Images\\
\hline
19. Q+Ka & Track 1 & I  & -0.25 & 0.051$\times$0.034 & 22.8 & 27 - 47 & 8.1 & \ref{continuum-multiepoch} \\
20. Q+Ka & Track 2 & I & -0.25 & 0.047$\times$0.035 & 24.9 & 27 - 48 & 8.0 & \ref{continuum-multiepoch}\\
21. Q+Ka & Track 3 & I & -0.25 & 0.048$\times$0.034 & 23.1 & 27 - 48 & 8.0 & \ref{continuum-multiepoch}\\
22. Q+Ka & Track 4 & I & -0.25 & 0.048$\times$0.036 & 21.8 & 27 - 48 & 8.0 & \ref{continuum-multiepoch}\\
\hline
2012 Data\\
\hline
23. Q & All Tracks & I & -0.5 & 0.058$\times$0.033 & 42.4 & 40 - 42 & 7.3 & \ref{continuum-2012}\\
\enddata
\tablecomments{Here we describe all the images used in the analysis presented here. Note that we do not
show all images in the paper for the sake of brevity, but we include their properties since they
were used to measure flux densities and/or analyzed the upper limits on polarized flux density.}
\tablenotetext{a}{Images are not shown in the paper, but they are used for the approximately
beam-matched flux density measurements of both sources combined in Table \ref{flux-densities} and Figure \ref{radio-spectrum}.}
\label{image-parameters}
\end{deluxetable}

\begin{deluxetable}{llllll}
\tabletypesize{\scriptsize}
\tablewidth{0pt}
\tablecaption{Flux densities}

\tablehead{\colhead{Wavelength} & \colhead{Flux Density} & Peak Intensity & Image Robust & \colhead{Reference} &\colhead{Table \ref{image-parameters} Image}\\
            \colhead{(mm)}      & \colhead{(mJy)} &    \colhead{(mJy~beam$^{-1}$)} & & &
}
\startdata
Combined\\
\hline
1.3 & 181$\pm$30 & 98.5 & \nodata & 3 & \nodata\\
3.4 & 9.0$\pm$1.0 & 6.8 & \nodata & 3 & \nodata\\
6.52 & 2.0$\pm$0.1 (1.7) & 0.51 & 2.0 & 1 & 15 \\
7.14 & 1.6$\pm$0.1 (1.3) & 0.47 & 1.5  & 1 & 16\\
7.3 & 1.7$\pm$0.1  & 0.49 & 2.0 & 2 & \nodata\\
8.45 (27-48~GHz) & 1.1$\pm$0.04 & 0.47 & 2.0 & 1& 1\\
8.1  & 1.0$\pm$0.05 (0.77) & 0.34 & 0.5  & 1 & 17\\
10.3 & 0.53$\pm$0.05 (0.30) & 0.23 & 0.0 & 1 & 18\\
14  & 0.54$\pm$0.07 & 0.31 & 2.0 & 2& \nodata\\
33  & 0.24$\pm$0.02 & 0.21 & 2.0 & 2& \nodata\\
36  & 0.21$\pm$0.07 & 0.21 & 2.0 & 2& \nodata\\
62  & 0.18$\pm$0.05 & 0.19 & 2.0 & 2& \nodata\\
\hline
A (western) source\\
\hline
6.52 &  0.33$\pm$0.05 & 0.33 & 0.0 & 1 & 11\\
7.14 &  0.31$\pm$0.03 & 0.31 & 0.0 & 1& 12\\
8.1  &  0.27$\pm$0.03 & 0.27  & -0.5 & 1 & 13\\
8.45 (27-48~GHz) & 0.27$\pm$0.03 & 0.27 & -0.25 & 1 & 2\\
10.3  & 0.27$\pm$0.02 & 0.27  & -0.5 & 1 & 14 \\
\hline
B (eastern) source\\
\hline
6.52  & 0.27$\pm$0.05 & 0.27 & 0.0 & 1 & 11\\
7.14 &  0.27$\pm$0.03 & 0.27 & 0.0 & 1 & 12\\
8.1   & 0.23$\pm$0.03 & 0.22 & -0.5 & 1& 13\\
8.45 (27-48~GHz) & 0.20$\pm$0.02 & 0.2 & -0.25 & 1 & 2\\ 
10.3 &  0.1$\pm$0.02 & 0.1 & -0.5 & 1 & 14\\
\enddata
\tablecomments{Values presented in this table are the result of Gaussian fitting to the source(s). The fits
to the A and B source flux densities have their major and minor axes fixed to the size of the synthesized
beam at each wavelength. The values in parentheses in the Flux
Density column have their estimated free-free contribution subtracted.
References: 1 - This work, 2 - \citep{tobin2013}, 3 - \citep{chiang2012}. The last column, Table \ref{image-parameters} 
\textbf{
Image refers to the image number in Table 3 that the flux densities were derived from.
}
}
\label{flux-densities}
\end{deluxetable}

\end{document}